\documentclass[aps,nofootinbib,prd,superscriptaddress,tightenlines,notitlepage,twocolumn,showpacs,floatfix]{revtex4-1}

\usepackage[colorlinks=true, citecolor=magenta]{hyperref}
\usepackage{tabularx}
\usepackage{bm}
\usepackage{graphicx}
\usepackage{amssymb,bm,tensor}
\usepackage{xcolor}
\usepackage{amsmath}
\usepackage[varg]{txfonts}
\usepackage{enumerate}
\usepackage[normalem]{ulem}
\usepackage{bm}
\usepackage[OT1]{fontenc}
\usepackage{float}

\usepackage{scalerel,stackengine}

\begin{document}

\title{Non-Equilibrium Relativistic Core Collapse of Self-Interacting Dark Matter Halos -- Limits On Seed Black Hole Mass
}

\author{Hua-Peng Gu}
\affiliation{Department of Astronomy, School of Physics, Peking University, Beijing 100871, China}
\affiliation{Kavli Institute for Astronomy and Astrophysics, Peking University, Beijing 100871, China}

\author{Fangzhou Jiang}
\email[Corresponding author: ]{fangzhou.jiang@pku.edu.cn}
\affiliation{Department of Astronomy, School of Physics, Peking University, Beijing 100871, China}
\affiliation{Kavli Institute for Astronomy and Astrophysics, Peking University, Beijing 100871, China}

\author{Xian Chen}
\email[Corresponding author: ]{xian.chen@pku.edu.cn}
\affiliation{Department of Astronomy, School of Physics, Peking University, Beijing 100871, China}
\affiliation{Kavli Institute for Astronomy and Astrophysics, Peking University, Beijing 100871, China}

\author{Ran Li}
\affiliation{School of Physics and Astronomy, Beijing Normal University, Beijing 100875, China}
\affiliation{School of Astronomy and Space Science, University of Chinese Academy of Sciences, Beijing 100049, China}

\date{\today}

\begin{abstract}
Recent observations of supermassive black holes (SMBHs) at high redshifts pose
challenges to standard seeding mechanisms. Among competing models, the collapse
of self-interacting dark matter (SIDM) halos provide a plausible explanation
for early SMBH formation. While previous studies on modeling the gravothermal
collapse of SIDM halos have primarily focused on non-relativistic evolution
under the assumption of hydrostatic equilibrium, We advance this framework by
relaxing the equilibrium assumption and additionally incorporating
general-relativistic effects. To this end, we introduce the Misner-Sharp
formalism to the SIDM context for the first time. Our model reproduces the
standard hydrostatic models in the early long-mean-free-path (LMFP) regime, but
displays interesting distinct behavior in the late short-mean-free-path (SMFP)
regime, where intense outward heat flux drives a rapid expansion of the outer
envelope, removing mass from the core and significantly decelerating the
collapse. Our general relativistic treatment enables us to follow halo
evolution to the final stage when the apparent horizon forms. Our simulation
yields a seed black hole mass of approximately $3\times10^{-8}$ of the halo
mass at horizon formation, suggesting that additional mechanisms such as
baryonic effects are critical for seeding black holes that are
sufficiently massive to account for SMBHs in the early Universe.
\end{abstract}

\maketitle

\section{\label{sec:Introduction}Introduction}

The observation of high-redshift, luminous quasars has revealed the existence
of supermassive black holes (SMBHs) in the early universe
\citep{SDSS:2001emm,Willott:2003xf,Fan:2022fhc}. Traditionally, these black
holes (BHs) are thought to originate from stellar-mass seeds—remnants of the
first massive stars \citep{Heger:2002by}, which grow through continuous
accretion \citep{Salpeter:1964kb,Hopkins:2009td,Inayoshi:2019fun} or mergers
\citep{Kauffmann:1999ce,Bhowmick:2024jwc}. However, the James Webb Space
Telescope (JWST \citep{Gardner:2006ky}) challenged this ``light-seed'' paradigm
by revealing a population of massive compact objects known as the ``little red
dots'' (LRDs) \citep{Matthee:2023utn} which have been interpreted by many
studies as actively accreting SMBHs
\citep{Greene:2024phl,2025ApJ...978...92L,Inayoshi:2025isg}. The LRDs appear
unexpectedly abundant at redshifts $4 < z < 9$ and are believed to host BHs
with masses ranging from $10^5$ to $10^7 \, \mathrm{M}_{\odot}$
\citep{2024ApJ...968...38K,2024arXiv240403576K,2026Natur.649..574R}. However, their stellar
masses are surprisingly low relative to the scaling relations observed in local
active galactic nuclei
(AGNs)\citep{Goulding:2023gqa,2024A&A...691A.145M,Chen:2025mzw}. Reaching such
masses by these redshifts requires a seed BH mass of $\gtrsim 10^4 \,
\mathrm{M}_{\odot}$ and uninterrupted Eddington accretion. Consequently, such
discovery indicates a significant tension in the standard seeding mechanism,
suggesting that light-seed scenarios are insufficient to explain the abundance
of high-redshift SMBHs \citep{Bogdan:2023ilu}.

To explain the origin of high-redshift SMBHs, various baryonic mechanisms have
been proposed, including population III remnants, gas-dynamical processes, and
stellar-dynamical processes. Population III remnants, however, typically yield
seeds of only $\sim 10^2 - 10^3 \, \mathrm{M}_{\odot}$
\citep{Volonteri:2010wz,Barack:2018yly,Chantavat:2023dfg}, insufficient to
become a promising candidate. More massive seeds may arise from gas-dynamical
processes, where the primordial gas cloud could directly collapse into a
compact, massive object under specific conditions, without fragmentation and
the subsequent formation of individual stars
\citep{Bromm:2002hb,Begelman:2006db}. Such BH seeds, also named as the
direct-collapse BHs (DCBHs), can exceed $10^4 \, \mathrm{M}_{\odot}$
\citep{Bromm:2002hb,Lodato:2006hw,Shang:2009ij}. Yet, this channel requires the
suppression of $\mathrm{H}_2$ cooling via intense UV flux to prevent
fragmentation, a condition difficult to achieve without the formation of
first-generation stars \citep{Latif:2013dua,Glover:2000ph}. Alternatively,
stellar-dynamical processes in dense clusters can produce $10^2 - 10^4 \,
\mathrm{M}_{\odot}$ seeds via run-away collisions
\citep{Devecchi:2012nw,Kritos:2024upo}. However, this mechanism demands extreme
stellar densities ($\gtrsim 10^7 \, \mathrm{M}_{\odot}/\mathrm{pc^3}$) and low
metallicity environments that rarely exist simultaneously
\citep{Latif:2016qau,Roberts:2024wup}.

Beyond heavy seeds, mechanisms accelerating BH growth, such as super-Eddington
accretion \citep{Santoro:2005tv,Inayoshi:2015pox} and mergers
\citep{Johnson:2012cw,Inayoshi:2019fun,Bhowmick:2024jwc}, have been explored.
While these can theoretically form SMBHs rapidly from light seeds, they face
significant environmental constraints. Sustained super-Eddington accretion
requires rare, high-density, low-velocity gas reservoirs \citep{Shi:2022hud},
while mergers are limited by gravitational-wave recoil kicks, which often eject
BHs from their fuel sources \citep{Haiman:2004ve,Yoo:2004ze,Inayoshi:2019fun}.
Moreover, some cosmological origins, including various types of primordial BHs (PBHs)
\citep{Khlopov:2008qy,Barack:2018yly,Escriva:2021aeh,Escriva:2022duf} are also
considered, where large density fluctuations in the early universe may decouple
from the cosmic expansion and collapse into BHs
\citep{Zeldovich:1967lct,Hawking:1971ei,Carr:1974nx}. The most stringent constraints on the mass spectrum of heavy PBHs come from the cosmic microwave background (CMB), which accounts for the fact that the radiation from gas accretion onto such massive PBHs would preionize the intergalactic medium and significantly distort the CMB angular power spectra
\citep{Carr:2018rid,Inayoshi:2019fun,2026arXiv260106024C}. While the
theoretical possibility exists \citep{Inomata:2016uip,Escriva:2022duf}, the
lack of observation renders the scenario overcomplex and requiring further
observations to verify.

Meanwhile, significant progress has been made in dark matter (DM) research.
Extensive astronomical observations \citep{Freeman:1970mx,Brainerd:1995da} have
established Cold Dark Matter (CDM) as the prevailing paradigm. In this model,
DM behaves as a collisionless fluid interacting exclusively via gravity
\citep{WMAP:2010qai,Planck:2018vyg}, forming large-scale structures and
virialized halos
\citep{1980lssu.book.....P,Davis:1985rj,2010gfe..book.....M,Vogelsberger:2019ynw}.
However, lacking radiative cooling or thermal conduction due to the
collisionless nature, CDM halos cannot collapse further to form BHs
\citep{Lynden-Bell:1966zjv,2010gfe..book.....M,Arguelles:2020qsi,Cirelli:2024ssz}.
Despite its success on large scales, the CDM paradigm faces challenges in matching some small-scale observations \citep{Bullock:2017xww}. Even by considering baryonic processes that modify CDM halos, the structural diversity of dwarf galaxies cannot be fully accommodated (see e.g., a review by \citet{2022NatAs...6..897S} and references therein). 

This motivates alternative models that modifies the collisionless nature of dark matter. The elastic self-interacting DM (SIDM) provides a particularly promising candidate \citep{Spergel:1999mh}. By introducing a finite self-interaction cross section $\sigma$ \citep{Tulin:2017ara,Correa:2020qam}, the SIDM model provides viable solutions to most of the small-scale discrepancies while preserving the large-scale success of CDM \citep{ Spergel:1999mh,Vogelsberger:2012ku,Rocha:2012jg,Peter:2012jh,Zavala:2012us,Oman:2015xda,Yang:2021kdf}. This is achieved because self-interactions thermalize the central halo, creating the necessary structural diversity—where profiles can be cored or cuspy depending on the cross-section as well as the formation time of the halo—to match observations. According to current observations, the cross section per unit particle mass is constrained to \citep{Gilman:2021sdr,Adhikari:2022sbh}$
\sigma \sim 1-100\ \mathrm{cm^2/g}$, with possible velocity dependence that makes the cross section decrease sharply beyond a relative velocity of $\sim 100 \, \mathrm{km/s}$ \citep{Correa:2020qam}. By heat conduction from the inner core to the outer region \citep{Spergel:1999mh}, these collisions eventually lead to a run-away gravothermal core collapse that occurs within several hundred relaxation timescales \citep{Balberg:2002ue,Balberg:2001qg}. At high redshifts, the core-collapse timescale in high-concentration and early-forming halos can be shorter or comparable to the age of the Universe \citep{Jiang:2025jtr}. Hence, this mechanism offers a plausible pathway for the formation of high-redshift SMBHs \citep{Arguelles:2023hab}.

The gravothermal collapse of SIDM halos has been extensively studied in a variety of previous works. Such a process has been modeled using a fluid approximation, where the halo is treated as a spherically symmetric, non-relativistic ideal fluid that undergoes quasi-static collapse under the assumption of hydrostatic equilibrium. This formalism was originally introduced for the study of globular star clusters \citep{Lightman:1978zz,Lynden-Bell:1980xip}, and was first applied to SIDM halos by \citet{Balberg:2002ue}. Since then, many studies have addressed the core-collapse time and the density-profile evolution using the fluid model, including the presence of baryonic components \citep{Feng:2020kxv,Zhong:2023yzk}, partially ultra-strong interactions \citep{Pollack:2014rja,Roberts:2024wup}, dissipative interactions \citep{Essig:2018pzq,Choquette:2018lvq,Shen:2021frv,Xiao:2021ftk,Shen:2025evo}, and velocity-dependent cross sections \citep{Outmezguine:2022bhq,Yang:2022zkd,Gad-Nasr:2023gvf}. Complementing the fluid approach, $N$-body simulations have also been conducted \citep{Koda:2011yb,Rocha:2012jg,Sameie:2018chj,Robertson:2018anx,Banerjee:2019bjp,Sabarish:2025hwb}, although they face significant computational challenges in resolving the core collapse at late times \citep{Palubski:2024ibb,Fischer:2024eaz}.

However, none of these previous studies have been able to determine the final
seed BH mass—a parameter essential for assessing the viability of SIDM
core-collapse as a high-redshift SMBH seeding mechanism. The limitations of the
fluid approximation stem from two fundamental shortcomings: (1) the lack of
incorporation of general relativity (GR), and (2) the
reliance on the unrealistic assumption of hydrostatic equilibrium.  In the
Newtonian framework, late-stage evolution leads to a ``gravothermal
catastrophe," where central density and temperature diverge
\citep{Balberg:2002ue,Lynden-Bell:1980xip}. However, this formalism breaks down
as the system inevitably reaches the relativistic limit. In contrast, GR
describes this dynamical evolution in a completely different way, where the
metric tensor evolves and eventually develops a singularity signaling the
formation of the horizon \citep{Penrose:1964wq,Hawking:1971vc}. This
necessitates a relativistic framework to determine the final BH mass.
Furthermore, the inclusion of GR alone is insufficient because the assumption of
hydrostatic equilibrium breaks down during late-stage evolution. Applying the general stability criterion derived in \citet{Feng:2021rst}, as the velocity dispersion becomes relativistic, the system's adiabatic index eventually drops below the rising critical threshold required for stability, thereby triggering a dynamical instability. This
transition from quasi-static evolution to rapid gravitational collapse occurs
before horizon formation, rendering the hydrostatic approximation invalid.
Therefore, currently there exist only estimations of the seed BH mass, based on
extrapolation \citep{Gad-Nasr:2023gvf} or energy conservation
\citep{Feng:2025rzf}. To rigorously determine this mass, a complete
time-dependent GR analysis tracing the non-equilibrium evolution towards
horizon formation is required.

Incorporating GR into the non-equilibrium evolution will introduce profound physical consequences that were previously overlooked. As matter falls deep into the relativistic potential well near the horizon, a substantial amount of gravitational energy—comparable to the rest mass energy—is released. In the case of elastic SIDM collisions, which lack radiative dissipation, this energy is converted into kinetic energy and effectively conducted outward as heat. This intense outward heat flux drastically increases the pressure in the outer layers, acting against the collapse. Consequently, this mechanism raises critical questions about the efficiency of subsequent accretion and whether a massive seed BH can indeed form. However, this dynamic effect is entirely missed in previous studies due to the strict enforcement of hydrostatic equilibrium. For example, existing general relativistic treatments [e.g., \citealp{Shapiro:2014oha,Shapiro:2018vju}] have often been restricted to static equilibrium solutions, thereby neglecting the consequences of dynamical gravitational energy release. Thus, resolving the true evolutionary trajectory requires capturing this interplay between relativistic collapse and the opposing thermal pressure, which is only possible through a fully non-equilibrium approach.

Calculating the non-equilibrium collapse of a spherically symmetric system is a well-established issue in GR. This calculation can be traced back to \citet{Oppenheimer:1939ue}. In their work, the collapse of a system composed of pressureless dust with uniform density is semi-analytically calculated. Furthermore, incorporating the pressure, as expressed through the equation of state (EoS), \citet{Misner:1964je} derived the Misner-Sharp equations, which were widely employed in calculations about primordial BHs \citep{Escriva:2022duf} and compact object collapse \citep{May:1966zz,1978ApJ...221..304V,1979ApJ...232..558V,1995ApJ...443..717B}. Notably, the Misner-Sharp equations provide an ideal tool for our purposes, as they encompass both GR effects and dynamical evolution beyond equilibrium. In this work, we apply this formalism to modeling SIDM halo collapse for the first time. However, since the standard equations are adiabatic, we must incorporate heat conduction to adapt the framework for our analysis. We achieve this by introducing a heat flux term into the energy-momentum tensor. This extended general relativistic framework enables us to trace the full evolutionary history of the halo—from the initial virialized state to the formation of the BH apparent horizon—providing a self-consistent solution to the SMBH seeding problem.

The paper is organized as follows. We begin in
Sec.~\ref{sec:Methods} by reviewing the traditional fluid approximation before
introducing our general relativistic formalism based on the Misner-Sharp
equations. In Sec.~\ref{sec:Results}, we present the full numerical evolution
of the SIDM halo, tracing the trajectory from the quasi-static LMFP core
expansion phase, through the dynamic SMFP collapse, to the final formation of
the BH horizon. Section \ref{sec:Discussion} addresses the implications of
these results for the SMBH seeding problem, discussing the potential mechanisms
that can affect the final mass. Finally, a summary is provided in
Sec.~\ref{sec:Conclusion}.

\section{\label{sec:Methods}Methods}

In this section, we introduce the theoretical framework employed to model the evolutionary trajectory of collapsing SIDM halos. We begin by reviewing the traditional fluid approximation before introducing our extended general relativistic formalism based on the Misner-Sharp equations. Throughout this work, we adopt natural units where $c=G=1$. 

\subsection{\label{sec:Methods1}Traditional fluid method}

Under the standard fluid approximation and the assumption of hydrostatic equilibrium, the gravothermal evolution of an SIDM halo is governed by the following set of equations \eqref{eq:gravothermal1} - \eqref{eq:gravothermal4} \citep{Balberg:2002ue}:
%--
\begin{equation}\label{eq:gravothermal1}
\frac{\partial M}{\partial r}=4 \pi r^2 \rho,
\end{equation}
%--
%--
\begin{equation}\label{eq:gravothermal2}
\frac{\partial P}{\partial r}=-\frac{M \rho}{r^2},
\end{equation}
%--
%--
\begin{equation}\label{eq:gravothermal3}
\frac{L}{4 \pi r^2}=-\kappa \frac{\partial T}{\partial r},
\end{equation}
%--
%--
\begin{equation}\label{eq:gravothermal4}
\frac{\partial L}{\partial r}=-4 \pi r^2 \rho\left\{\left(\frac{\partial}{\partial t}\right)_M \frac{3 v^2}{2}+P\left(\frac{\partial}{\partial t}\right)_M \frac{1}{\rho}\right\}.
\end{equation}
%--
Here, the system is characterized by the matter density $\rho(r,t)$, enclosed mass $M(r,t)$, pressure $P(r,t)$, temperature $T(r,t)$, velocity dispersion $v(r,t)$, and luminosity $L(r,t)$. These equations correspond to mass conservation \eqref{eq:gravothermal1}, hydrostatic equilibrium \eqref{eq:gravothermal2}, thermal conduction \eqref{eq:gravothermal3}, and energy conservation \eqref{eq:gravothermal4}, respectively. In thermal equilibrium, the temperature is given by $kT=mv^2$ and the EoS is 
%--
    \begin{equation}\label{eq:EoS}
P=\rho v^2.
\end{equation}
%--

The conductivity $\kappa$ in Eq.\eqref{eq:gravothermal3} exhibits different behaviors based on the ratio of the mean free path $\lambda=1/(\rho \sigma)$ to the scale height that corresponds to the local dynamical time $H=\sqrt{v^2/(4\pi\rho)}$. In the short mean free path (SMFP) regime, where $\lambda \ll H$, the conductivity follows the standard form $\kappa_{\text{SMFP}} = (3/2)b\rho\lambda^2/(amt_0)$. Here, the constants are $a=\sqrt{16/\pi}$ and $b = 25\sqrt{\pi}/32$ \citep{1970mtnu.book.....C}, while $t_0=\lambda/(av)$ is the relaxation time. Conversely, in the long mean free path (LMFP) regime ($\lambda \gg H$), heat transport is described by $\kappa_{\text{LMFP}} = (3/2)C\rho H^2/(mt_0)$ \citep{Lynden-Bell:1980xip}, where $C$ serves as a calibration parameter. Following \citet{Koda:2011yb}, we adopt $C=0.75$, a value that enables the fluid evolution to match $N$-body simulation results. To smoothly interpolate between these two limits, the effective conductivity is constructed as $\kappa^{-1} = \kappa_{\text{SMFP}}^{-1} + \kappa_{\text{LMFP}}^{-1}$ \citep{Balberg:2002ue}. In the LMFP regime, heat flux is directly proportional to the collision rate, so that $\kappa \propto \sigma$; while in the SMFP regime, where frequent collisions hinder transport, $\kappa \propto \sigma^{-1}$.

\subsection{\label{sec:Methods2}Misner-Sharp formalism}

To capture the full non-equilibrium dynamics and general relativistic effects, we must extend the static framework to include bulk motion and spacetime curvature. Following \citet{Misner:1964je}, we adopt a time-dependent, spherically symmetric metric in the general form:
%--
\begin{equation}\label{eq:metric-misner}
ds^2=-e^{2\phi(r,t)}dt^2+e^{\lambda(r,t)} dr^2+R(r,t)^2d\Omega^2,
\end{equation}
%--
where $R(r,t)$ represents the circumferential radius. We adopt a Lagrangian (comoving) coordinate system where the radial coordinate $r$ labels specific fluid shells. In this frame, the fluid 4-velocity simplifies to $u^\mu=(e^{-\phi},0,0,0)$. It is convenient to utilize the enclosed rest mass $A$ as the radial coordinate. Since the rest mass enclosed within a comoving shell is conserved \footnote{In SIDM model, if the collision between DM particles is sufficient, then mass shell crossing is avoided.}, $A$ is a strictly monotonically increasing function of $r$:
%--
\begin{equation}\label{eq:Ar}
A(r) = 4 \pi \int_0^r \rho e^{\lambda / 2} R^2 d r,
\end{equation}
%--
where $\rho$ is the rest mass density. By defining the radial coordinate directly as the rest mass (i.e., choosing $r=A$), the metric component $e^{\lambda}$ is constrained by the rest mass conservation:
%--
\begin{equation}\label{eq:elambda}
e^{-\lambda/2}=4\pi R^2\rho.
\end{equation}
%--
Consequently, the metric in mass coordinates becomes:
%--
\begin{equation}\label{eq:metric-misnerA}
ds^2=-e^{2\phi(A,t)}dt^2+e^{\lambda(A,t)} dA^2+R(A,t)^2d\Omega^2.
\end{equation}
%--

To describe the gravothermal evolution of the SIDM halo, we extend the perfect fluid energy-momentum tensor to including a heat flux term:
%--
\begin{equation}\label{eq:energy-momentum-q}
    T^{\mu\nu} = (\rho' + P) u^\mu u^\nu + P g^{\mu\nu} + q^\mu u^\nu + q^\nu u^\mu,
\end{equation}
%--
where $\rho'=(1+\epsilon)\rho$ is the total energy density, $\epsilon$ is the specific internal energy, $P$ is the pressure, and $q^\mu$ represents the heat flux 4-vector. We adopt Eckart's formulation for relativistic thermodynamics, where the purely conductive heat transfer is given by \citep{Eckart:1940te,Misner:1973prb}:
%--
\begin{equation}\label{eq:Eckart}
q^\mu = -\kappa h^{\mu\nu} \left( \nabla_\nu T + T a_\nu \right),
\end{equation}
%--
with the projection tensor $h^{\mu\nu}\equiv g^{\mu\nu}+u^\mu u^\nu$ and 4-acceleration $a_\nu \equiv u^\alpha \nabla_\alpha u_\nu$. The orthogonality condition $q^\mu u_\mu = 0$, combined with spherical symmetry, implies that $q^t=0$ and the angular components vanish. Thus, only the radial component $q^r$ remains non-zero, taking the specific form \citep{Shapiro:2018vju}:
%--
\begin{equation}\label{eq:qr}
    q^r = -\kappa e^{-\lambda} e^{-\phi} \frac{\partial}{\partial r}\left(T e^{\phi}\right).
\end{equation}
%--

We further introduce the following fundamental quantities defined in the Misner-Sharp formalism.

\noindent The Misner-Sharp mass, representing the total mass-energy enclosed within shell $A$:
%--
\begin{equation}\label{eq:mr}
m(A,t)\equiv4 \pi \int_0^A \rho(1+\epsilon) R^2(\partial R / \partial A) d A.
\end{equation}
%--
\noindent The generalized Lorentz factor:
%--
\begin{equation}\label{eq:Gamma}
\Gamma \equiv e^{-\lambda / 2} \frac{\partial R}{\partial A}.
\end{equation}
%--
\noindent The coordinate velocity, describing the bulk radial motion:
%--
\begin{equation}\label{eq:ms1}
U\equiv e^{-\phi} \frac{\partial R}{\partial t}.
\end{equation}
%--
\noindent Finally, the specific enthalpy:
%--
\begin{equation}\label{eq:ms2}
w\equiv1+\epsilon+\frac{P}{\rho}.
\end{equation}
%--
By substituting the metric and energy-momentum tensor into the Einstein field equations and the conservation laws ($\nabla_\mu T^{\mu\nu}=0$), we derive the complete set of hydrodynamic evolution equations \citep{Misner:1964je,Liebendoerfer:2000fw}:
%--
\begin{equation}\label{eq:ms3}
\frac{\partial m}{\partial A}=(1+\epsilon)\Gamma+\frac{Uq}{\rho},
\end{equation}
%--
%--
\begin{equation}\label{eq:ms4}
\Gamma=\left(1+U^2-\frac{2m}{R}\right)^{1/2},
\end{equation}
%--
%--
\begin{equation}\label{eq:ms5}
   \frac{\partial\phi}{\partial A}=-\frac{1}{w} \left[\frac{1}{\rho} \frac{\partial P}{\partial A}+e^{-\phi}\frac{\partial}{\partial t}\left( \frac{q}{4\pi R^2 \rho^2}  \right)  \right],
\end{equation}
%--
%--
\begin{equation}\label{eq:ms6}
\rho=\frac{\Gamma}{4\pi R^2(\partial{R}/\partial{A})},
\end{equation}
%--
%--
\begin{equation}\label{eq:ms7}
\frac{\partial \epsilon}{\partial t}=-P \frac{\partial}{\partial t}\left(\frac{1}{\rho}\right)-e^{-\phi}\frac{\partial}{\partial A}\left(4 \pi R^2 qe^{2\phi}\right),
\end{equation}
%--
%--
\begin{equation}\label{eq:ms8}
\frac{\partial U}{\partial t}  =-e^{\phi}\left[-e^{-\phi}\Gamma^2  \frac{\partial \phi}{\partial A}\left(\frac{\partial R}{\partial A}\right)^{-1}+\frac{m}{R^2}+4 \pi P R\right].
\end{equation}
%--
To close the system, we adopt an ideal gas EoS:
%--
\begin{equation}\label{eq:ms9}
P=(\gamma-1)\epsilon\rho,
\end{equation}
%--
where $\gamma$ is the adiabatic index. For the non-relativistic case, $\gamma=5/3$, recovering the standard relation Eq. \eqref{eq:EoS} via $\epsilon=\frac{3}{2}kT$. Furthermore, by expressing temperature in terms of specific internal energy $\epsilon$ and utilizing Eq. \eqref{eq:elambda} to eliminate $\lambda$, the heat flux equation \eqref{eq:qr} can be rewritten in mass coordinates as:
%--
\begin{equation}\label{eq:ms10}
q = -(\gamma-1)\kappa e^{-\phi} \Gamma(4\pi R^2\rho) \frac{\partial}{\partial A}\left( e^{\phi}\epsilon\right).
\end{equation}
%--

Equations \eqref{eq:ms1} - \eqref{eq:ms10} constitute the generalized Misner-Sharp formalism, extended here to include the heat conduction term $q$. This system allows us to solve for the time evolution of the ten dynamical quantities ($R$, $U$, $m$, $\rho$, $P$, $\epsilon$, $w$, $\Gamma$, $\phi$, and $q$) given appropriate initial and boundary conditions. Notably, Eq. \eqref{eq:ms8} serves as the acceleration equation, replacing the hydrostatic equilibrium condition \eqref{eq:gravothermal2}. By transforming Eq. \eqref{eq:ms8} back to the spatial coordinate $r$ and taking the adiabatic limit ($q=0$), we recover the generalization of the TOV equation \citep{Tolman:1939jz,Oppenheimer:1939ne} to the non-static case:
%--
\begin{equation}\label{eq:TOVplus}
e^{-\phi}\frac{\partial U}{\partial t}=\Gamma e^{-\lambda/2}\frac{\partial\phi}{\partial r}-\frac{m+4 \pi R^3 P}{R^2}.
\end{equation}
%--
This form explicitly demonstrates how the bulk acceleration $\partial U/\partial t$ is driven by the competition between the pressure gradient (first term) and the effective gravity (second term, which includes relativistic corrections). Equation \eqref{eq:ms7} represents energy conservation, replacing Eq. \eqref{eq:gravothermal4}, while Eq. \eqref{eq:ms10} serves as the relativistic heat conduction law, replacing Eq. \eqref{eq:gravothermal3}.

\subsection{\label{sec:Methods3}Horizon formation}

A critical phenomenon in the gravitational collapse of compact objects is the emergence of horizons. To investigate this, we analyze the behavior of outgoing radial null geodesics during the collapse. Based on the metric in Eq. \eqref{eq:metric-misnerA}, the derivative operator along such a geodesic is defined as \citep{Hernandez:1966zia}:
%--
\begin{equation}\label{eq:null}
D_k\equiv e^{-\phi}\frac{\partial}{\partial t}+e^{-\lambda/2}\frac{\partial}{\partial A}.
\end{equation}
%--
It follows immediately that $D_kA=e^{-\lambda/2}>0$. This indicates that the rest mass coordinate $A$ increases along the null geodesic, implying that the light ray propagates outward relative to the local matter fluid. However, this does not guarantee that the photon physically expands to larger radial distances. If the matter collapses rapidly enough, these light rays may be trapped. To quantify this, we combine the definition of $\Gamma$ in Eq. \eqref{eq:Gamma} with the Misner-Sharp equations \eqref{eq:ms1} and \eqref{eq:ms4}, yielding the evolution of the areal radius along the ray:
%--
\begin{equation}\label{eq:horizon}
D_k R=\frac{1-2 m R^{-1}}{\Gamma-U}.
\end{equation}
%--

For any event $(t,A)$ satisfying the condition $R(t,A)<2m(t,A)$, then it can be deduced that $D_kR(t,A)<0$, since $\Gamma$ remains positive in this framework, and for infalling matter $U<0$. Consequently, light emitted from such a region is dragged inward and cannot escape to infinity. The condition $R(t,A)=2m(t,A)$ then defines a trapped surface, also named as the apparent horizon, within which even outgoing null geodesics are confined \citep{Penrose:1964wq}. In contrast, defining the event horizon requires knowledge of the global causal structure of spacetime. During gravitational collapse, the apparent horizon, if it exists at a given time slice, must lie within or coincide with the event horizon \citep{Hawking:1973uf}. In summary, this Misner-Sharp formalism provides a self-consistent method to precisely calculate the seed BH mass at the moment of apparent horizon formation in the SIDM halo collapse scenario.

\subsection{\label{sec:Methods4}Numerical setup \& method validation}

Using the Misner-Sharp formalism, we can numerically solve for the evolution of SIDM halos. We initialize the system using a Navarro-Frenk-White (NFW) density profile \citep{Navarro:1995iw,Navarro:1996gj}:
%--
\begin{equation}\label{eq:nfw}
\rho(R,t=0)=\frac{\rho_\mathrm{s}}{\left(R/{R_\mathrm{s}}\right)\left(1+R/{R_\mathrm{s}}\right)^{2}}.
\end{equation}
%--
We adopt typical halo parameters $R_\mathrm{s}=2.6\,\mathrm{kpc}$ and $\rho_\mathrm{s}=0.019\,\mathrm{M}_{\odot}/\mathrm{pc^3}$ \citep{Nishikawa:2019lsc}, corresponding to a halo mass of $M(<10R_\mathrm{s})=6.3\times10^{9}\,\mathrm{M}_{\odot}$. With a cross-section of $\sigma=5\,\mathrm{cm}^2 / \mathrm{g}$, the characteristic relaxation timescale, defined at the scale radius $R=R_\mathrm{s}$, is $t_0=0.26\,\mathrm{Gyr}$. To facilitate the relativistic analysis, we employ natural units and normalize all physical quantities by the total halo mass $M$. These parameters constitute our fiducial model for the subsequent analysis.

The numerical solution is performed using a Lagrangian discretization scheme based on the mass coordinate $A$. The system is partitioned into concentric spherical shells, where each shell is labeled by a fixed enclosed rest mass $A$. Unlike Eulerian grids, the mass coordinate $A$ remains constant for each shell, while its radius $R(A,t)$, bulk velocity $U(A,t)$, and thermodynamic variables (e.g., $\epsilon(A,t)$) evolve dynamically. For the boundary conditions, at the center ($A=0$), we impose the regularity conditions $R=0$, $U=0$, $m=0$, and $\Gamma=1$ \citep{1995ApJ...443..717B}. At the outer boundary ($A=A_{\mathrm{total}}$), we set $e^\phi=1$\footnote{This implies that we adopt the proper time of a comoving observer at the outer boundary as our coordinate time. Theoretically, the NFW profile extends to infinity; however, for numerical feasibility, we truncate the domain at $R=20R_\mathrm{s}$ ($A_{\mathrm{total}}=1.405M$). Since the halo matter continues beyond this radius, the pressure at the truncation boundary is non-zero, rather than a vacuum condition.}. With these initial and boundary conditions, we numerically solve the coupled system of Eqs. \eqref{eq:ms1} - \eqref{eq:ms10} using a finite-difference algorithm. Further details on the numerical implementation are provided in Appendix~\ref{sec:app1}.

\begin{figure}
    \centering 
    \includegraphics[width=1.08\linewidth]{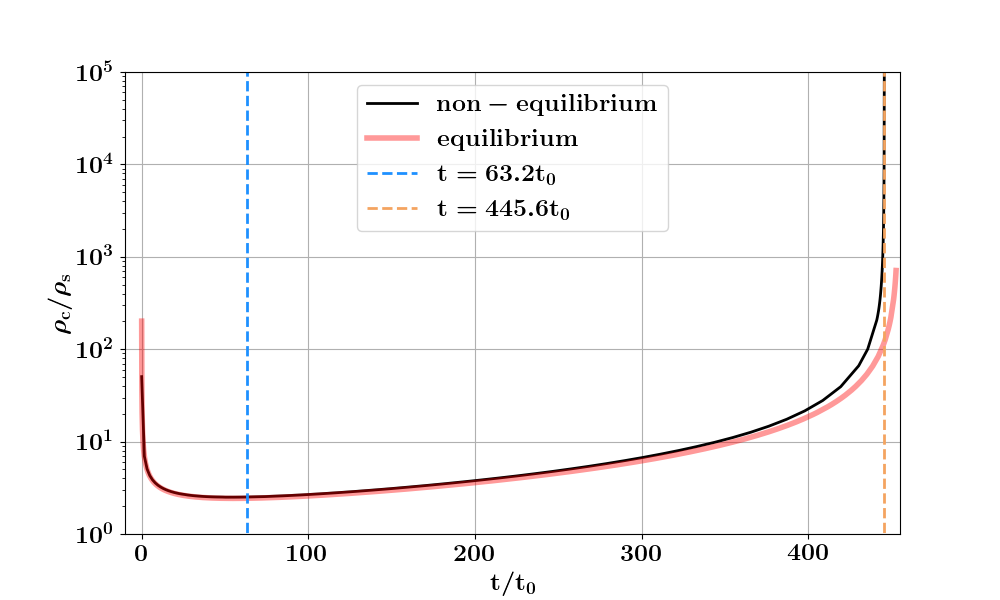} 
\caption{\label{fig:fig1} The evolution of the central density with time, comparing the traditional method (assuming hydrostatic equilibrium, red line) with our proposed method (non-equilibrium, black line). Our non-equilibrium treatment agrees well with the equilibrium fluid model during core formation and the early stages of core collapse, but predicts faster collapse in the late stages.
} 
\end{figure}

To validate our formalism and quantify the impact of non-equilibrium effects, we performed a parallel calculation using the traditional hydrostatic method governed by Eqs. \eqref{eq:gravothermal1} - \eqref{eq:gravothermal4}, assuming the same initial conditions. Figure \ref{fig:fig1} compares the central density evolution derived from both approaches. The excellent agreement observed in the early stage confirms the validity of our method in this quasi-static regime. However, as the system evolves, significant deviations appear. In the late stages, the high central density generates a massive heat flux that drives the system away from equilibrium, causing the quasi-static approximation to break down. In this phase, our framework reveals a faster collapse. Physically, this acceleration occurs because the bulk velocity accumulates dynamically via the acceleration equation \eqref{eq:ms8}, whereas the traditional method artificially resets this bulk motion to zero at each time step to enforce hydrostatic equilibrium. Employing our non-equilibrium formalism is essential to correctly describe the late-time evolutionary stage and to precisely calculate the final BH mass.

\section{\label{sec:Results}Results -- \NoCaseChange{Evolution towards BH formation}}

\begin{figure*}[t]
    \centering
    \makebox[\textwidth][c]{
    \hspace{0.5cm}
    \includegraphics[width=0.57\textwidth]{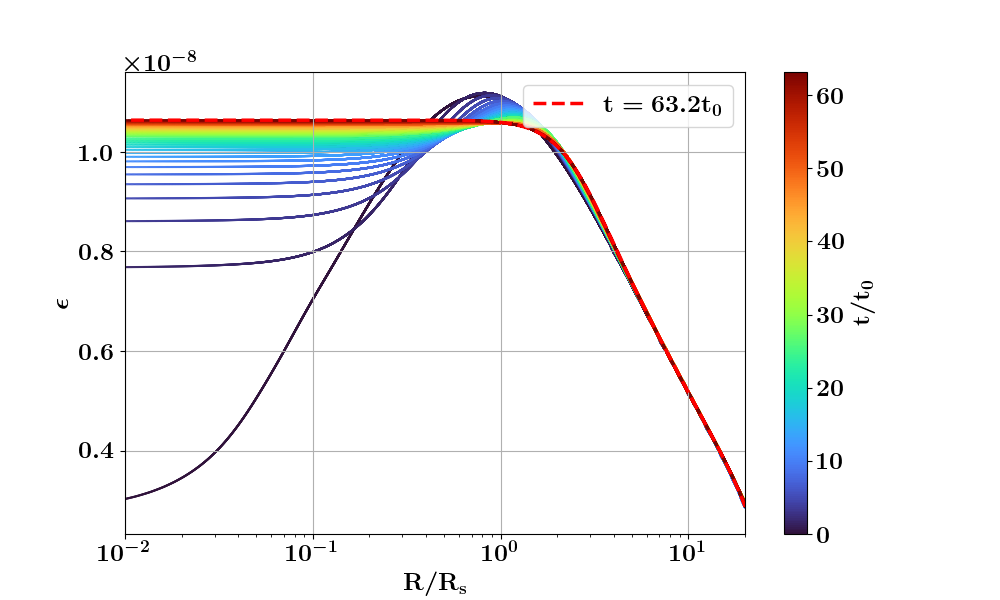} 
    \hspace{-1cm}
    \includegraphics[width=0.57\textwidth]{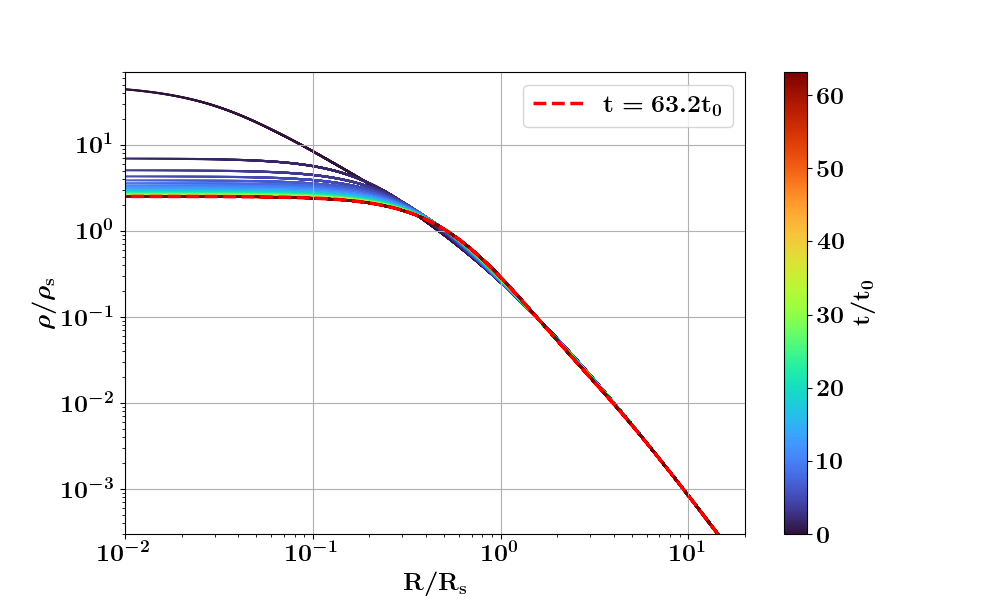}
    }
    \makebox[\textwidth][c]{
    \hspace{0.5cm}
    \includegraphics[width=0.57\textwidth]{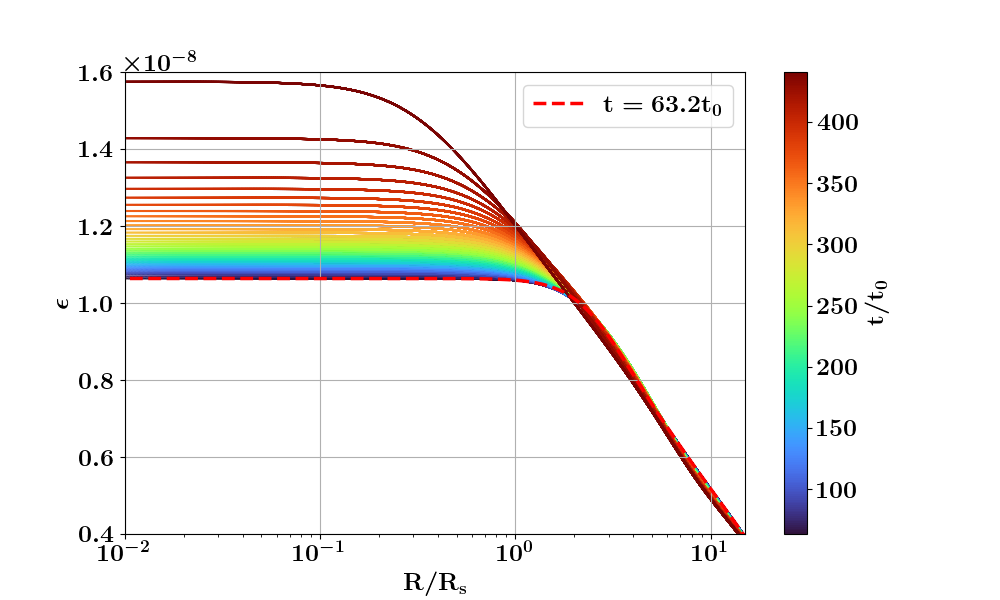}
    \hspace{-1cm}
    \includegraphics[width=0.57\textwidth]{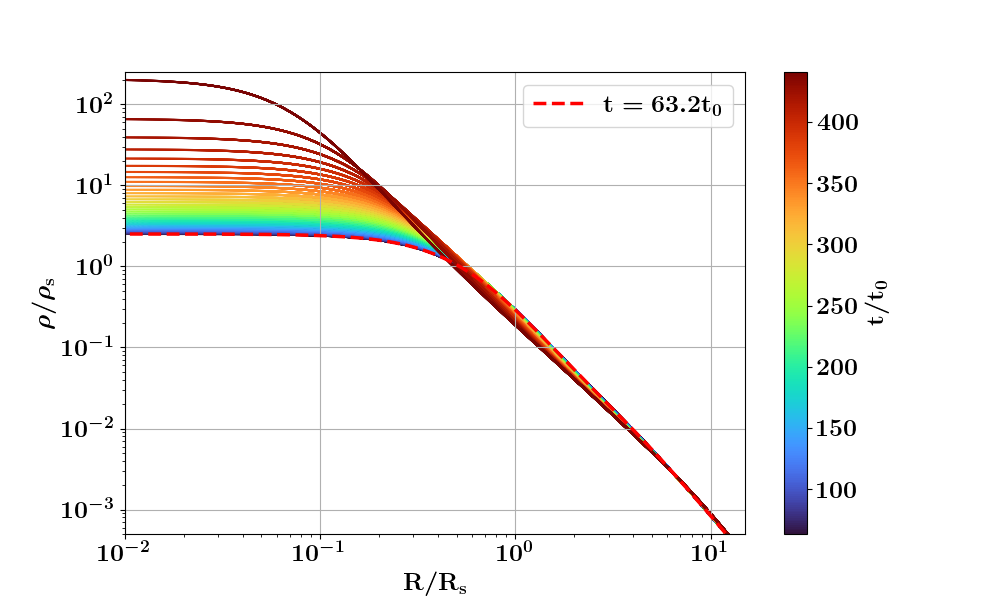}
    }
    \caption{\label{fig:fig2} Evolution of the specific internal energy (left column) and density profiles (right column) for the fiducial halo. The upper panels illustrate the core expansion phase, where the halo evolves from an initial NFW profile ($t=0$) towards an isothermal core. The lower panels depict the subsequent gravothermal collapse phase. The red dashed line at $t=63.2t_0$ marks the formation of the isothermal core, serving as the transition boundary between the expansion and collapse regimes. The color gradient of the solid lines represents the progression of time. Note that $\epsilon$ denotes the specific internal energy normalized by $c^2$}.
 
\end{figure*}

In this section, we present the complete evolutionary trajectory of the SIDM halo, tracing the pathway from an initial NFW profile to the eventual formation of a BH. This continuous process naturally decomposes into three stages corresponding to distinct dynamical regimes. First, in Sec.~\ref{sec:Results1}, we examine the early gravothermal evolution, tracing the core expansion and self-similar collapse in the LMFP regime. Next, in Sec.~\ref{sec:Results2}, we investigate the SMFP phase, where outward heat flux and non-equilibrium effects significantly reshape the core structure. Finally, in Sec.~\ref{sec:Results3}, we track the terminal relativistic collapse leading to the formation of the BH horizon.

\subsection{\label{sec:Results1}Gravothermal evolution}

The first stage of SIDM halo evolution is core expansion, a phase extensively discussed in previous literature \citep{Ahn:2004xt,deBlok:2009sp,DiCintio:2017zdz,Kaplinghat:2019dhn}. The upper panels of Figure \ref{fig:fig2} illustrate the evolutionary trajectories of the specific internal energy and density profiles in this stage, calculated using our Misner-Sharp formalism. Initially, the NFW halo configuration is in hydrostatic equilibrium at all radii, resulting in zero bulk velocity throughout the system. However, heat conduction due to SIDM collisions redistributes the specific internal energy, modifying the pressure gradient. Consequently, the pressure force no longer balances gravity, inducing bulk motion. Given the temperature profile of an NFW halo, heat is initially transported primarily from radius $R \sim R_\mathrm{s}$ toward the center. This process heats the central region, transforming the initial density cusp into an expanding isothermal core. Eventually, a strictly isothermal core forms at the center by $t=63.2t_0$. It is important to note, however, that the physical expansion phase concludes earlier, at $t=51.3t_0$ when the central density reaches its minimum. Subsequently, the core density begins to rise as the energy loss from outward heat transfer exceeds the gain from inward flow. By $t=63.2t_0$, the inward heat flux vanishes entirely, leaving only outward flux, which marks the definitive entry into the gravothermal collapse phase. At the moment of maximum expansion, the central density is $ \rho(t=51.3t_0,A=0)=2.5\rho_\mathrm{s}$.

The evolution of the specific internal energy and density during the subsequent gravothermal collapse stage is shown in the lower panels of Figure \ref{fig:fig2}. In this phase, heat is transferred from the core region ($R \gtrsim R_\mathrm{s}$) to the outer halo, causing the core to contract. As the core shrinks and its density increases, the heat flux intensifies, accelerating the collapse. For moderate central densities ($\lesssim 10^2\rho_\mathrm{s}$), the LMFP condition $\lambda \gg H$ holds globally, leading to a self-similar evolution \citep{Balberg:2002ue}\footnote{Strictly speaking, self-similarity assumes a quasi-static evolution. In our formalism, which includes bulk velocity, the evolution approximates self-similarity in the LMFP regime because the bulk velocity remains much smaller than the velocity dispersion.}. This self-similar collapse persists until $t \approx 445t_0$, when the LMFP condition breaks down in the core region. Before this time, the bulk velocity $U$ is negligible compared to the velocity dispersion $v$\footnote{In our relativistic formalism, we utilize specific internal energy $\epsilon$ rather than $v$. These quantities are related via $v^2=(\gamma-1)\epsilon$.}.

\subsection{\label{sec:Results2}SMFP evolution}

\begin{figure}
    \centering 
    \includegraphics[width=1.1\linewidth]{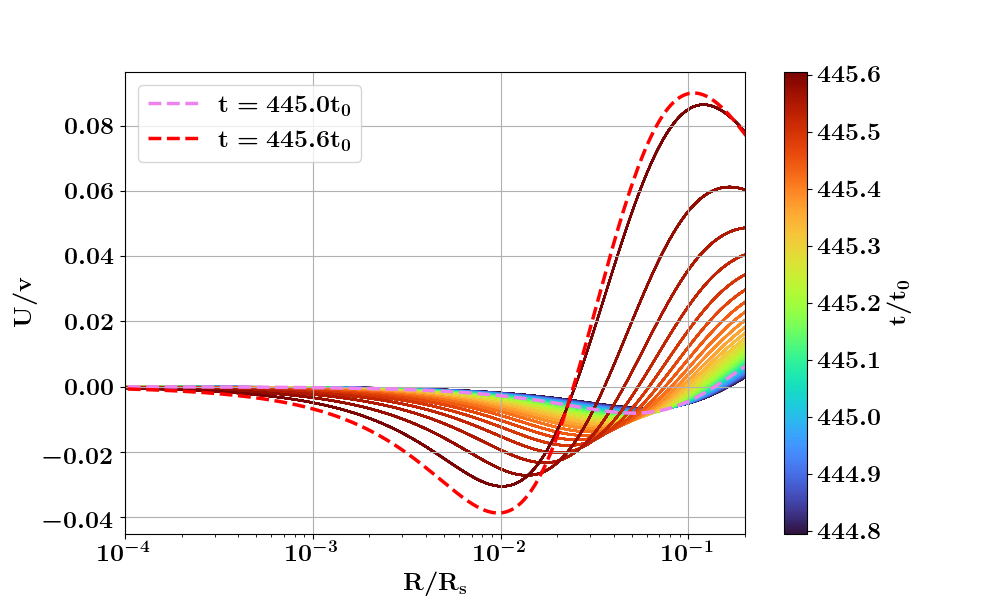} 
    \vspace{1em}
    \includegraphics[width=1.1\linewidth]{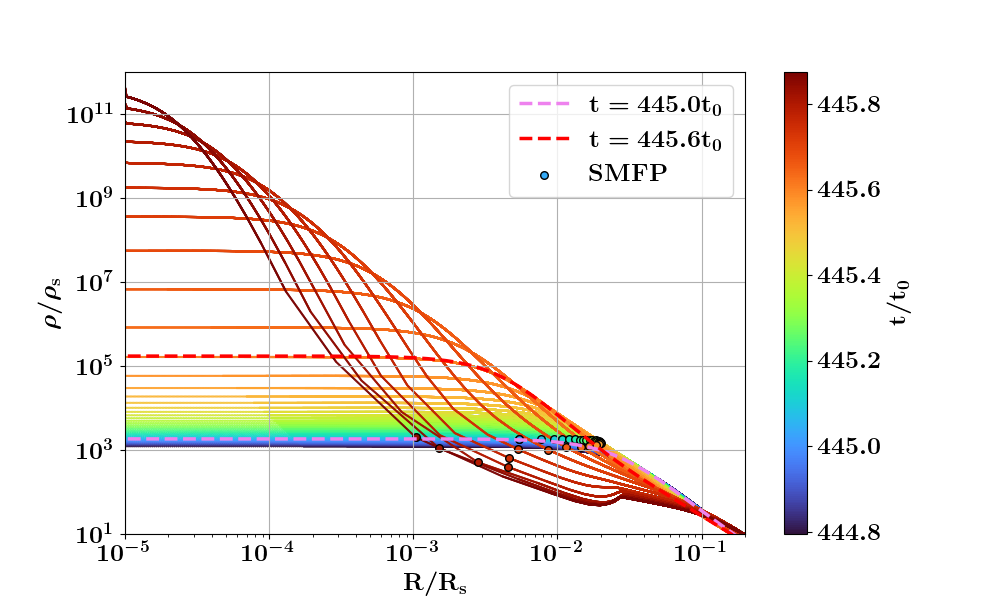}
\caption{\label{fig:fig3} The evolution of the bulk velocity profile (upper panel) and the density profile (lower panel) during the SMFP core collapse stage. The notations follow those in Figure \ref{fig:fig2}. In the upper panel, the y-axis represents the ratio of the bulk velocity $U$ to the velocity dispersion $v$, which serves as a diagnostic for deviations from hydrostatic equilibrium. The black-edged dots mark the boundary of the SMFP regime at each evolutionary snapshot. The density depletion in the outer region arises from mass outflow, which is induced by the intense outward heat flux characteristic of the SMFP regime.
} 
\end{figure}

In our fiducial model, non-equilibrium effects become dominant once the core enters the SMFP regime, corresponding to central densities $\rho_c \gtrsim 10^5\rho_\mathrm{s}$. The evolution of the bulk velocity profile during this stage is presented in the upper panel of Figure \ref{fig:fig3}. The system exhibits a distinct bifurcation into two dynamic regions: a collapsing inner core and a rapidly expanding outer envelope. The interface between these regions propagates inward, indicating that the core is simultaneously shrinking in radius and losing mass due to outflow. This strong mass outflow is driven by the intense heat flux emerging from the high-density core, where thermal conduction is highly efficient. This energy is absorbed by the shells surrounding the core, triggering a rapid expansion with velocities reaching a significant fraction of the velocity dispersion. The resulting density evolution, shown in the lower panel of Figure \ref{fig:fig3}, reveals that this rapid expansion creates a significant density gap just outside the core. Consequently, non-equilibrium dynamics fundamentally alter the density profile during the SMFP phase.

\begin{figure}
    \centering 
    \includegraphics[width=1.08\linewidth]{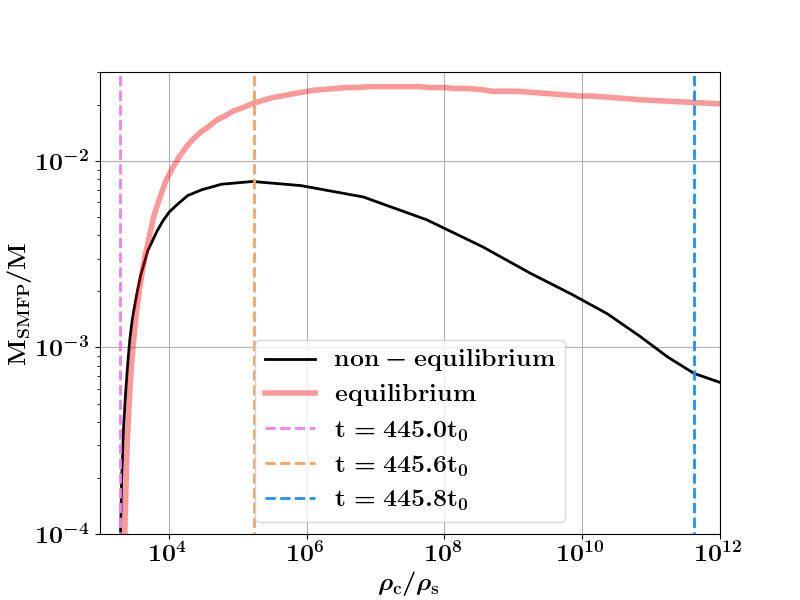} 

\caption{ \label{fig:fig4} Evolution of the SMFP mass fraction with central density, comparing the traditional method (assuming hydrostatic equilibrium, red line) with our proposed method (non-equilibrium, black line). In the non-equilibrium case, the SMFP regime emerges at $t=445.0t_0$, reaches its maximum enclosed mass at $t=445.6t_0$, and gradually stabilizes after $t=445.8t_0$. In contrast, the equilibrium calculation yields a substantially larger SMFP mass.}

\end{figure}

A critical parameter in SMFP evolution is the enclosed mass within the SMFP regime ($M_{\text{SMFP}}$), as it provides the fundamental mass budget for the seed BH. Previous estimates generally posit that a small fraction of $M_{\text{SMFP}}$ eventually collapses into the central BH \citep{Balberg:2002ue,Nishikawa:2019lsc,Meshveliani:2022rih,Gad-Nasr:2023gvf, Feng:2025rzf}. The boundary of the SMFP regime is indicated by the black-edged dots in the lower panel of Figure \ref{fig:fig3}.To clearly show the evolution of the SMFP mass, we present the trajectory of the SMFP mass fraction in Figure \ref{fig:fig4}, indicated by the black line. The transition begins at $t=445.0t_0$, when the central density rises sufficiently to satisfy the $\lambda=H$ condition. Subsequently, $M_{\text{SMFP}}$ grows, reaching a maximum of approximately $5\times10^7 \, \mathrm{M}_{\odot}$ ($1\%$ of the total halo mass) at $t=445.6t_0$. Following the peak at $t=445.6t_0$, the SMFP mass begins to decline. This reversal occurs because the core continuously loses mass to outflow, while the surrounding material—already depleted by the aforementioned density gap—is too diffuse to effectively replenish the SMFP region. Shortly thereafter, by $t \approx 445.8t_0$, the density profile exterior to the core stabilizes\footnote{Stability here refers to the density profile shape; it does not imply that mass outflow has ceased. Mass outflow persists, but since the remaining core mass is negligible compared to the total mass of the outer shells, further outflow does not significantly perturb the outer density structure.}, and a dense core with $\rho > 10^{10}\rho_\mathrm{s}$ forms within the region $R < 10^{-3}R_\mathrm{s}$. The SMFP mass declines to $\sim10^{-4}$ of the total halo mass at $t = 445.8t_0$ and gradually plateaus afterwards, reflecting the stabilization of the outer halo. 

Non-equilibrium effects play a pivotal role here, as they substantially alter the density profile. Our calculated maximum SMFP mass is slightly lower than that reported by \citet{Nishikawa:2019lsc}. This discrepancy arises because our non-equilibrium framework accounts for dynamical accumulation of outflow velocity due to continuous heat conduction, whereas quasi-static approaches determine the outflow rate solely based on the instantaneous thermal state. To quantify this deviation, we plot the hydrostatic result in Figure \ref{fig:fig4} (red line) for comparison. In the hydrostatic case, the SMFP mass fraction peaks at $\sim 2.5\%$ and nearly stabilizes at this value in the subsequent evolution. This implies that the unrealistic assumption of the hydrostatic equilibrium would overestimate the SMFP mass, which is critical for determining the final BH mass.

\subsection{\label{sec:Results3}Black hole formation}

By the end of SMFP evolution, although the central density of the SMFP core eventually exceeds $\rho > 10^{10}\rho_\mathrm{s}$, the compactness $m/R$ within the core grows only to $\sim 10^{-6}$ in natural units, remaining far from the BH formation threshold. In the late-time evolution, the core mass decreases due to continuous mass outflow, yet the core radius shrinks at a faster rate, leading to an increase in the ratio $m/R$ over time. Our general relativistic formalism enables us to track the evolution continuously until the horizon formation ($R=2m$), thereby allowing us to directly calculate the seed BH mass at formation, which has never been exactly realized in previous studies.

\begin{figure}
    \centering 
    \includegraphics[width=1.1\linewidth]{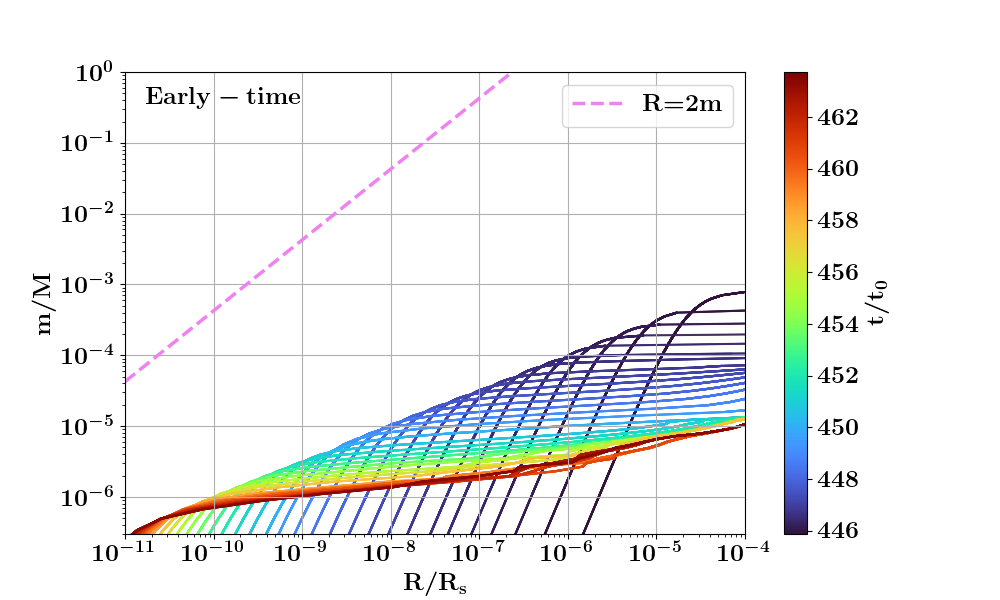} 
    \vspace{1em}
    \includegraphics[width=1.1\linewidth]{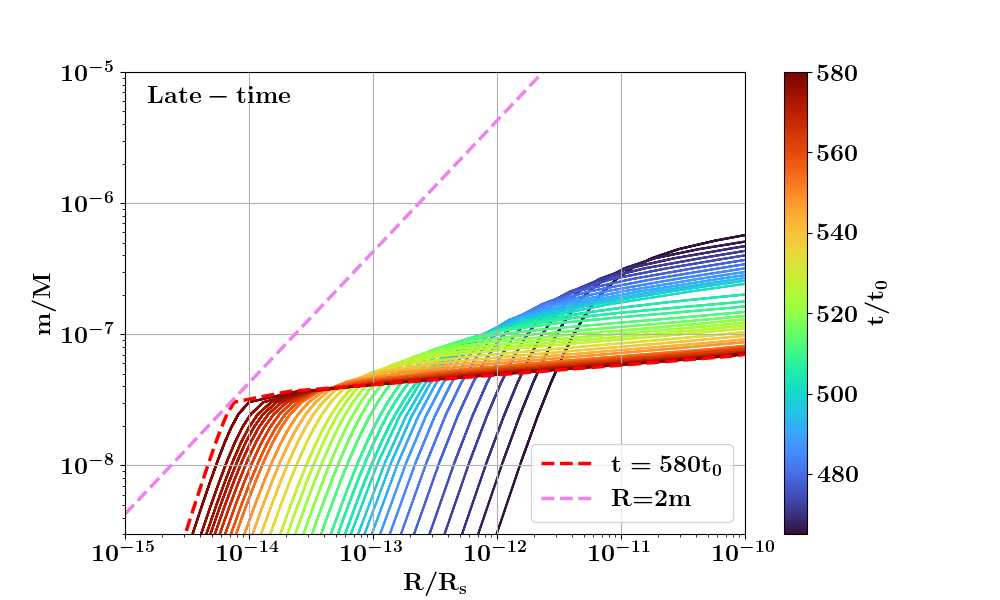}
\caption{\label{fig:fig5} The evolution of the mass profile at BH formation stage. The upper panel traces the $m-R$ evolution in the non-relativistic regime ($m_\mathrm{core}/R_\mathrm{core} \lesssim 10^{-2}$), while the lower panel displays the strong-gravity regime approaching the horizon ($m_\mathrm{core}/R_\mathrm{core} \gtrsim 10^{-2}$). The violet dashed lines represent the apparent horizon condition $R=2m$. At $t=580t_0$, the mass profile (red dashed line) intersects $R=2m$, marking the formation of the central BH. Other notations follow those in Figure \ref{fig:fig2}. The seed BH mass at horizon formation is about $3\times10^{-8}$ of the halo mass ($\approx 200\ \mathrm{M}_{\odot}$).
} 
\end{figure}

To visualize the system's approach towards horizon formation, we present the evolution of the mass-radius ($m-R$) relation in Figure \ref{fig:fig5}. Within these profiles, the core radius $R_\mathrm{core}$ corresponds to the prominent turning point, marking the boundary of the collapsing core. In the non-relativistic regime (upper panel, where $m_\mathrm{core}/R_\mathrm{core} \ll 1$), the collapse exhibits self-similar behavior again. While both the core mass and radius decrease continuously, the shape of the mass profile remains invariant. The envelope of the mass profiles approximates a power-law relation $m_{\mathrm{core}} \propto R_{\mathrm{core}}^{1/2}$, indicating an increase in compactness over time. This scaling follows from mass conservation during self-similar evolution, where $d( \ln m_{\text{core}} ) / d( \ln R_{\text{core}} ) = 3 - \alpha$ and $\alpha$ is the power-law index for the tail of the density profile ( $\rho \propto R^{ - \alpha }$ ). While the classic LMFP regime predicts $\alpha \approx 2.2$ \citep{Lynden-Bell:1980xip}, our finding of $\alpha \approx 2.5$ reflects a tendency towards the SMFP regime in our conductivity model. It is important to note that, despite introducing bulk velocity into our formalism, the core's evolution in this regime can still be approximated as quasi-static. This is because the bulk velocity in the deep core remains negligible compared to the local sound speed. According to the acceleration equation \eqref{eq:ms8}, when the core shrinks to extremely small scales ($R_\mathrm{core} \sim 10^{-10}R_\mathrm{s}$), any slight deviation from hydrostatic equilibrium induces a massive acceleration via the pressure gradient, which immediately restores the system to its equilibrium state. Consequently, non-equilibrium effects are significant only in the rapidly expanding outer envelope, which have negligible impact on the central core dynamics. Leveraging this physical property, we adopt a quasi-static approximation while retaining GR effects for the central region during this phase to accelerate numerical calculation\footnote{In principle, the time step is strictly limited by the sound speed via the Courant-Friedrichs-Lewy condition (Eq. \eqref{eq:cfl}, see Appendix~\ref{sec:app1}). However, since the system's evolution is governed by the bulk velocity, virtually no physical evolution occurs within such a constrained time step, leading to extreme computational inefficiency. By adopting the quasi-static assumption, the instability associated with the sound speed is suppressed, allowing us to safely adopt a much larger time step.}.

Notably, Figure \ref{fig:fig5} reveals a counter-intuitive feature: the collapse rate of the core decreases over time, as evidenced by the extended duration required for the system to bridge the final orders of magnitude between $m$ and $R$. Such deceleration happens at $t \approx 445.8t_0$ due to rapid expansion of the surrounding shells, which stands in contrast to the standard expectation of a run-away collapse process \citep{Balberg:2002ue}. However, this phenomenon is physically consistent within our framework: the intense outward heat flux significantly increases the pressure in the outer layers, effectively impeding further collapse. We can verify this deceleration quantitatively by estimating the collapse timescale. In this late stage, the core evolution is governed by the conductive heat loss. Therefore, the characteristic collapse time can be estimated via
%--
\begin{equation}\label{eq:tc}t_\mathrm{c}=\frac{E_\mathrm{core}}{4\pi R_\mathrm{core}^2q_\mathrm{core}},
\end{equation}
%--
where $E_\mathrm{core}$ is the total internal energy of the core, and $q_\mathrm{core}$ is the heat flux at the core boundary. At $t = 445.8t_0$, when the core enters the self-similar collapse phase, we find $t_\mathrm{c}\sim0.1t_0$, signifying a rapid collapse. However, during the self-similar evolution, the core shrinks significantly, causing the total heat luminosity $4\pi R_\mathrm{core}^2q_\mathrm{core}$ to decrease. This reduction in energy loss rate progressively lengthens the collapse timescale $t_\mathrm{c}$. By $t \approx 460t_0$, the timescale increases to $t_\mathrm{c}\sim40t_0$, indicating that the collapse has become significantly slower.

As the compactness parameter grows toward the strong gravity regime ($2m/R \to 1$), the evolutionary trajectory begins to deviate from the self-similar solution, as illustrated in the lower panel of Figure \ref{fig:fig5}. During this phase, although the quasi-static approximation remains valid for the core, the general relativistic terms in Eq. \eqref{eq:ms8} significantly modify the hydrostatic equilibrium structure. The $\Gamma^2$ factor associated with the pressure term reduces the effective pressure support as $\Gamma\to0$ when $2m/R \to 1$, leading to a denser core. Consequently, the evolutionary track deviates from the $m_\mathrm{core} \propto R_\mathrm{core}^{1/2}$ scaling, exhibiting a noticeably shallower gradient in the logarithmic plot. This deviation implies a more rapid convergence towards $R=2m$ condition. When the central temperature rises sufficiently such that the velocity dispersion becomes comparable to the speed of light, the dynamical instability is triggered \citep{Feng:2021rst}, leading to a rapid dynamical collapse. Finally, at $t=580t_0$, the mass profile intersects the horizon condition $R=2m$ at $m\approx3\times10^{-8}M$. This event marks the formation of a central black hole with a mass of $M_{\text{BH}} \approx 200\ \mathrm{M}_{\odot}$.

\section{\label{sec:Discussion}Discussion}

Our numerical simulations, based on the general relativistic Misner-Sharp formalism, show that the fiducial SIDM halo collapses into a BH of $M_{\text{BH}} \approx 200$ $M_{\odot}$ at $t = 580 t_0$, corresponding to a collapse fraction of $3 \times 10^{ - 8 }$. This result aligns well with recent estimations of $\sim 10^{ - 8 }$ derived from self-similar extrapolation \citep{Gad-Nasr:2023gvf} and energy conservation arguments \citep{Feng:2025rzf}. Furthermore, according to the analytical framework in \citep{Ralegankar:2024zjd}, the theoretical upper bound of the mass ratio $m_{\text{BH}} / M$ should be $\sim 10^{ - 6 }$ if the fiducial parameters in our model are used. Our finding of $3 \times 10^{ - 8 }$ is consistent with these limits, reinforcing the physical validity and robustness of our dynamical simulations.

By providing the first fully relativistic
calculation of this BH formation process, we confirmed that some earlier works
\citep{Balberg:2002ue,Koda:2011yb,Nishikawa:2019lsc,Meshveliani:2022rih} have
overestimated the seed mass. Crucially, our complete evolutionary history
uncovers the physical origin of this discrepancy: previous studies
underestimated the non-equilibrium effects, where intense outward heat flux
from the core can effectively hinder the gravitational collapse.

Nevertheless, the resulting seed BH mass is significantly below the threshold ($\gtrsim10^4 \mathrm{M}_{\odot}$) typically required to explain the progenitors of observed high-redshift SMBHs. This suggests that the pure SIDM collapse scenario, under standard halo parameters, may face difficulties in directly producing sufficiently heavy seeds. Hence, in this section, we investigate the dependence of the seed BH mass on halo parameters and explore additional physical mechanisms—specifically the inclusion of baryons, post-formation accretion, or the velocity-dependent cross-section—that could be incorporated into our framework to enhance the seed mass. Additionally, as our current simulation truncates the outer halo during core contraction, the sensitivity of our results to this boundary treatment will be addressed in future work.

The relatively low mass of the seed BH is primarily a consequence of the prolonged mass outflow phase identified in Section \ref{sec:Results2}. As the core enters the SMFP regime, efficient outward heat conduction drives a significant portion of the core mass into the expanding envelope. Throughout this evolution, the core maintains a self-similar scaling relation of approximately $m_\mathrm{core} \propto R_\mathrm{core}^{1/2}$. This power-law dependence implies that the final seed mass is intrinsically coupled to the halo's initial structural parameters; specifically, the initial compactness $M/R_\mathrm{s}$ sets the baseline for this evolutionary trajectory, with higher compactness halos naturally yielding more massive seeds. Furthermore, the self-interaction cross-section $\sigma$ plays a critical role during the LMFP-SMFP transition. Since the thermal conductivity in the SMFP regime scales as $\kappa \propto \sigma^{-1}$, a larger $\sigma$ reduces the efficiency of heat conduction. This suppression of outward heat flux mitigates mass loss, allowing the SMFP core to retain a larger fraction of its mass and effectively elevating the normalization of the $m_\mathrm{core}-R_\mathrm{core}$ relation. To quantitatively investigate the dependence of the seed BH mass on these parameters, we performed two additional simulations with different parameter sets, evolving the system from the initial NFW profile to the formation of the SMFP core. The results are shown in Table~\ref{tab:t1}, where we also list the maximum SMFP mass in these simulations as they serve as the fundamental mass budget for the seed BH. 

\begin{table}
    \centering
    \caption{Dependence of the seed BH mass on model parameters. The columns list the halo mass, the characteristic radius of the initial NFW profile, the concentration, the self-interaction cross-section, the maximum SMFP mass, and the final BH mass\footnote{The latter two simulations were terminated upon the formation of the maximum SMFP core. Consequently, their final BH masses are estimated based on the self-similar scaling relation.}.}
    \label{tab:t1}
    \renewcommand{\arraystretch}{1.3} 
    \setlength{\tabcolsep}{3.4pt}       
    \begin{tabular}{cccccc}
        \hline\hline
        $M\,(M_\odot)$ & $R_\mathrm{s}\,(\mathrm{kpc})$ & $c$ & $\sigma\,(\mathrm{cm}^2/\mathrm{g})$ & Max $M_{\mathrm{SMFP}}\,(M_\odot)$ & $M_{\mathrm{BH}}\,(M_\odot)$ \\
        \hline
        $6.3\times10^{9}$ & $2.6$  & $15.8$ & $5$  & $5\times10^{7}$   & 200 \\
        $6.3\times10^{9}$ & $0.65$ & $76.7$ & $5$  & $1.7\times10^{8}$ & 700 \\
        $6.3\times10^{9}$ & $2.6$  & $15.8$ & $20$ & $1.2\times10^{8}$ & 500 \\
        \hline\hline
    \end{tabular}
\end{table}

A critical factor omitted in this DM-only simulation is the presence of baryons. In a realistic protogalactic halo, baryons would fall into the deep gravitational potential well created by the collapsing SIDM core. Unlike dark matter, baryons can undergo dissipative cooling via radiative processes, allowing them to condense more efficiently than the conductive SIDM fluid. The high-density SIDM core calculated in our work ($\rho > 10^{10}\rho_s$) would serve as an ideal gravitational trap for such infalling gas. Previous studies have demonstrated that the dynamical interplay between SIDM gravothermal collapse and baryonic components can significantly accelerate the collapse rate \citep{Sameie:2018chj,Robertson:2018anx,Feng:2020kxv,Zhong:2023yzk,bosch2025dynamicscoresselfinteractingdark}. Consequently, it is reasonable to infer that the resulting BH mass would increase correspondingly due to the deepened gravitational potential. 

Furthermore, a BH seed of baryonic origin, such as a Population III stellar remnant, may form at the halo center prior to the completion of the SIDM collapse. Although such a seed might be initially low-mass, its presence would fundamentally alter the system's evolution. In our current SIDM-only simulations, the BH emerges only at the very final stage; prior to this, despite extreme densities, the core mass remains insufficient to trigger horizon formation, and the dynamics are dominated by mass outflow. However, the existence of a pre-existing central BH imposes an absorptive inner boundary condition. This would effectively circumvent the mass outflow phase: matter infalling toward the center would be accreted onto the BH rather than accumulating pressure to drive an outward flow. Since mass accretion across the horizon is irreversible—matter cannot be ``pushed back" once accreted—this mechanism would prevent mass loss and facilitate rapid BH growth. We intend to incorporate these baryonic physics and central accretion dynamics in our future work.

It is important to emphasize that $M_{\text{BH}}\approx 200\ \mathrm{M}_{\odot}$ represents only the initial mass at the instant of horizon formation. Although our simulations indicate that at the time of horizon formation, accretion is inhibited by a steep pressure gradient driven by outward heat flux, this suppression may be temporary. Once the horizon forms, the core is trapped and the outward heat flux is causally cut off, potentially removing the thermal barrier and allowing the infall of outer shells to resume. Moreover, recent studies propose that the presence of baryons could further facilitate this accretion process \citep{Feng:2025rzf}, potentially leading to a much larger final mass. In the presence of baryons, a higher fraction of the halo mass may enter the SMFP regime, thereby offering a more substantial reservoir for seed BH accretion.

Based on our simulation results, we perform a simplified estimation of the potential growth of the seed BH via Bondi accretion \citep{Bondi:1952ni,2006MNRAS.365..345H}. At the instant of horizon formation, the typical sound speed in the outer regions is $c_\mathrm{s}\sim2000\, \mathrm{km/s}$, yielding a Bondi radius of $R_\mathrm{B} = 2GM_\mathrm{BH}/c_\mathrm{s}^2 \sim 1.5\times10^{-10}R_\mathrm{s}$. The enclosed mass within this radius is merely $\sim10^{-7}$ of the total halo mass ($\approx 600\ \mathrm{M}_{\odot}$). Although this instantaneous reservoir is small, the critical factor is the continuous inflow of material from the outer shells. Assuming steady-state accretion, the Bondi rate is \citep{Feng:2025rzf} $\dot{M}_\mathrm{B}=4\pi\lambda_s G^2M_\mathrm{BH}^2 \rho c_\mathrm{s}^{-3} \sim 10^8\ \mathrm{M}_{\odot}/\mathrm{Gyr}$, where $\lambda_s=0.25$ and we evaluate $\rho$ and $c_\mathrm{s}$ at $R_\mathrm{B}$. However, given the steep radial decline of the density profile, such an extreme accretion rate is unsustainable. To rigorously quantify the cumulative mass growth of the BH seed, it requires future extended simulations using our Misner-Sharp formalism.

Distinct from accretion dynamics, a velocity-dependent cross-section could also fundamentally alter the late-time evolution. As investigated by previous works [e.g., \citealp{Correa:2020qam}], the cross-section $\sigma$ is expected to be as high as $\sigma \gtrsim 100\,\mathrm{cm}^2/\mathrm{g}$ at low velocities but suppressed at $v\gtrsim10^3\,\mathrm{km/s}$. If this model were incorporated into our simulation, the high cross-section in the low-velocity outer halo would significantly shorten the relaxation time, accelerating the initial core formation. In our fiducial model, the BH formation time $580 t_0$ exceeds the cosmic age; thus, introducing a velocity-dependent cross-section may be essential\footnote{Alternatively, a higher halo concentration could also reduce the relaxation timescale $t_0$}. However, in the late stages as the core temperature rises, $\sigma$ would drop significantly. One possibility is that the core transitions into the LMFP regime where the collision rate becomes insufficient to sustain heat conduction, potentially causing the collapse to stall. Alternatively, if the core remains in the SMFP regime, the reduced cross-section would actually enhance thermal conductivity since $\kappa \propto \sigma^{-1}$, leading to a vastly different collapse trajectory. Resolving which of these competing mechanisms dominates requires future simulations incorporating velocity-dependent interactions.

In our numerical simulations we did not consider specific particle
nature for the SIDM. However, if 
SIDM particles are fermions, the Fermi degeneracy pressure imposes a
lower threshold for the mass of the BH to form via core collapse.
For example, according to Ref.~\citep{Arguelles:2023kqw,Feng:2025ybf},
given the mass $m_{\text{DM}}$ of a DM particle, the minimum
BH mass is
$m_{\text{BH}} \gtrsim 10^5M_{\odot}\ ( 1 \text{ MeV} / m_{\text{DM}} )^2$. 
Consequently, for an SIDM halo to successfully collapse into a BH, its mass must satisfy $M \gtrsim 3 \times 10^{ 12 } \ ( 1 \text{ MeV} / m_{\text{DM}} )^2 \ M_{\odot}$,
since we found 
$m_{\text{BH}} / M \approx 3 \times 10^{ - 8 }$ 
in our SIDM simulations (assuming a halo concentration of  $c \sim 15$).
This example indicates that the correlation between seed BH mass and halo mass 
may help us constrain the physical properties of SIDM.
\section{Conclusion}\label{sec:Conclusion}

In this work, we have developed a fully general relativistic and hydrodynamical framework based on the Misner-Sharp formalism to investigate the gravothermal collapse of SIDM halos. Unlike previous quasi-static approaches, our method explicitly captures non-equilibrium dynamic effects, enabling us to track the system's evolution continuously from the initial core expansion phase to the formation of an apparent horizon. 

Our numerical results demonstrate consistency with traditional quasi-static approaches during the early evolutionary stages, verifying the validity of our formalism. However, as the system becomes denser, our simulations reveal that the late-stage dynamics in the SMFP regime are dominated by intense outward heat flux. This thermal transport drives a substantial mass outflow, decelerating the collapse and depleting the core mass significantly before the BH forms. For a representative halo with $M \sim 10^9 \mathrm{M}_{\odot}$ and constant cross-section $\sigma=5\, \mathrm{cm}^2/\mathrm{g}$, the maximum SMFP mass is about $1\%$ of the halo mass and the resulting seed BH mass is approximately $200\, \mathrm{M}_{\odot}$—corresponding to a seed-to-halo mass ratio of $\approx 3\times 10^{-8}$. We also investigated their dependence on halo parameters and the cross-section of the SIDM.

This result suggests that pure SIDM gravothermal collapse with constant cross-section faces challenges in directly producing the heavy seeds required for high-redshift SMBHs. To reconcile this discrepancy, we propose that additional physical mechanisms are essential. Specifically, the inclusion of baryons—which facilitate accretion by deepening the potential—as well as the consideration of post-horizon evolution and velocity-dependent cross-section, is critical for enhancing the seed mass. Future extensions of our formalism incorporating these effects will be necessary to fully constrain the role of SIDM in SMBH formation. 

The code used to model SIDM halo evolution in the Misner-Sharp framework is available in Ref.~\cite{SIDMcode} and can be accessed at \href{https://github.com/Hua-Peng-G/SIDM}{https://github.com/Hua-Peng-G/SIDM}.

\begin{acknowledgments}
We thank Wei-Xiang Feng for helpful discussions. This work was supported by the National Key Research and Development Program of China (Grant No.2024YFC2207300) and the National Natural Science Foundation of China (Grant No. 12473037). FJ acknowledges support by the National Natural Science Foundation of China (NSFC, 12473007) and China Manned Space Program with grant no. CMS-CSST-2025-A03. RL acknowledges the support of National Nature Science Foundation of China (No 11988101), the science research grants from the China Manned Space Program (No. CMS-CSST-2025-A0), CAS Project for Young Scientists in Basic Research (No. YSBR-062).
\end{acknowledgments}

\appendix{\label{sec:app}}

\section{\label{sec:app1}Numerical Implementation}

We employ a finite-difference algorithm with a staggered Lagrangian grid to numerically solve the system of Eqs. \eqref{eq:ms1} - \eqref{eq:ms10}. We denote the temporal step by the superscript $n$ ($n=0,1,\dots$) and the spatial grid index by the subscript $i$. The spatial domain is discretized into $N$ mass shells, with grid interfaces defined by the enclosed mass $\{A_i\}_{i=0}^{N}$.

In this staggered grid scheme, variables are defined either at the cell interfaces or at the cell centers:
\begin{itemize}
    \item Interface variables (defined at $i$): radius $R$, bulk velocity $U$, Lorentz factor $\Gamma$, Misner-Sharp mass $m$, lapse function $e^\phi$, and heat flux $q$.
    \item Cell-centered variables (defined at $i-1/2$): Rest mass density $\rho$, specific internal energy $\epsilon$, pressure $P$, and specific enthalpy $w$.
\end{itemize}
The discretization proceeds sequentially as follows. Note that for variables defined at cell centers, values at the interface $i$ are approximated via arithmetic averaging, e.g., $\bar{P}_i \equiv (P_{i-1/2} + P_{i+1/2})/2$.

\noindent \textbf{1. Update Velocity ($U$):}
Using variables from the previous time step $n-1$:
%--
\begin{equation}\label{eq:msn1}
\begin{split}
U^n_i = & U^{n-1}_i + \Delta t \Bigg[ (\Gamma^{n-1}_i)^2 \left(\frac{\partial \phi}{\partial A}\right)^{n-1}_i \\
&\times \left( 4\pi (R^{n-1}_i)^2 \bar{\rho}^{n-1}_i \right) \\
&- (e^\phi)^{n-1}_i \left( \frac{m^{n-1}_i}{(R^{n-1}_i)^2} + 4\pi R^{n-1}_i \bar{P}^{n-1}_i \right) \Bigg],
\end{split}
\end{equation}
%--
where the potential gradient term is given by:
%--
\begin{equation}\label{eq:dphidA_num}
\begin{split}
\left(\frac{\partial \phi}{\partial A}\right)^{n-1}_i &= \frac{-1}{2\bar{w}^{n-1}_i} \Bigg[ \frac{P^{n-1}_{i+1/2} - P^{n-1}_{i-1/2}}{\bar{\rho}^{n-1}_i ( A_{i+1}/2 - A_{i-1}/2 )} \\
&+ \frac{\sigma}{(e^\phi)^{n-1}_i \Delta t} \Bigg( \frac{q^{n-1}_i}{4\pi (R^{n-1}_i)^2 (\bar{\rho}^{n-1}_i)^2} \\
&- \frac{q^{n-2}_i}{4\pi (R^{n-2}_i)^2 (\bar{\rho}^{n-2}_i)^2} \Bigg) \Bigg].
\end{split}
\end{equation}
%--

\noindent \textbf{2. Update Radius ($R$):}
%--
\begin{equation}\label{eq:msn2}
R^n_i = R^{n-1}_i + (e^\phi)^{n-1}_i U^n_i \Delta t.
\end{equation}
%--

\noindent \textbf{3. Update Lorentz Factor ($\Gamma$):}
We utilize $m^{n-1}_i$ as an approximation for $m^n_i$ to decouple the implicit dependence\footnote{Strictly, $\Gamma^n$ depends on $m^n$. However, calculating $m^n$ requires $\Gamma^n$. For sufficiently small time steps, using $m^{n-1}$ provides a valid first-order approximation.}:
%--
\begin{equation}\label{eq:msn3}
\Gamma^n_i = \sqrt{1 + (U^n_i)^2 - \frac{2 m^{n-1}_i}{R^n_i}}.
\end{equation}
%--

\noindent \textbf{4. Update Density ($\rho$):}
%--
\begin{equation}\label{eq:msn4}
\rho^n_{i-1/2} = \frac{ (A_i - A_{i-1}) (\Gamma^n_{i-1} + \Gamma^n_i)/2 }{ \frac{4\pi}{3} \left[ (R^n_i)^3 - (R^n_{i-1})^3 \right] }.
\end{equation}
%--

\noindent \textbf{5. Update Internal Energy ($\epsilon$):}
%--
\begin{equation}\label{eq:msn5}
\begin{split}
\epsilon^n_{i-1/2} &= \epsilon^{n-1}_{i-1/2} - P^{n-1}_{i-1/2} \left( \frac{1}{\rho^n_{i-1/2}} - \frac{1}{\rho^{n-1}_{i-1/2}} \right) \\
&- \frac{4\pi \Delta t}{\bar{e^\phi}^n_{i-1/2}} \left[ \frac{ \mathcal{F}^{n-1}_i - \mathcal{F}^{n-1}_{i-1} }{ A_i - A_{i-1} } \right],
\end{split}
\end{equation}
%--
where $\mathcal{F}_i \equiv (R_i)^2 q_i (e^\phi_i)^2$.

\noindent \textbf{6. Update Pressure ($P$) and Enthalpy ($w$):}
%--
\begin{equation}\label{eq:msn6}
P^n_{i-1/2} = (\gamma - 1) \rho^n_{i-1/2} \epsilon^n_{i-1/2},
\end{equation}
%--
%--
\begin{equation}\label{eq:msn7}
w^n_{i-1/2} = 1 + \epsilon^n_{i-1/2} + \frac{P^n_{i-1/2}}{\rho^n_{i-1/2}}.
\end{equation}
%--

\noindent \textbf{7. Update Metric Potential ($\phi$):}
Integrated inward from the boundary condition $(e^\phi)^n_N=1$:
%--
\begin{equation}\label{eq:msn8}
\begin{split}
\ln((e^\phi)^n_i) &= \ln((e^\phi)^n_{i+1}) \\
&- \frac{1}{2} (A_{i+1} - A_i) \left[ \left(\frac{\partial \phi}{\partial A}\right)^{n}_i + \left(\frac{\partial \phi}{\partial A}\right)^{n}_{i+1} \right].
\end{split}
\end{equation}
%--

\noindent \textbf{8. Update Heat Flux ($q$):}
%--
\begin{equation}\label{eq:msn9}
\begin{split}
q^n_i = & -\frac{3}{2}(\gamma-1)^{3/2}a \times \left( \frac{1}{C} + \frac{a}{b} \frac{\sigma^2}{4\pi} \bar{P}^n_i \right)^{-1} \\
&\times \frac{\Gamma^n_i\sqrt{\bar{\epsilon}^n_i} \cdot \bar{P}^n_i \cdot (R^n_i)^2  \cdot 2 \left[ (\epsilon e^\phi)^n_{i+1/2} - (\epsilon e^\phi)^n_{i-1/2} \right] }{ (e^\phi)^n_i \cdot \bar{\rho^{-1}}^n_i \cdot (A_{i+1} - A_{i-1}) },
\end{split}
\end{equation}
%--
where $\bar{\rho^{-1}}$ denotes the average of the inverse density.

\noindent \textbf{9. Update Misner-Sharp Mass ($m$):}
%--
\begin{equation}\label{eq:msn10}
\begin{split}
m^n_i = & m^n_{i-1} + (A_i - A_{i-1}) \Bigg[ \bar{\Gamma}^n_i (1 + \epsilon^{n-1}_{i-1/2}) \\
&+ \frac{ (U^n_{i-1} + U^n_i) (q^n_{i-1} + q^n_i) }{ 4 \rho^{n-1}_{i-1/2} } \Bigg],
\end{split}
\end{equation}
%--
with the boundary condition $m^n_0=0$.

The numerical time step $\Delta t$ is strictly constrained by the Courant-Friedrichs-Lewy (CFL) condition \citep{1967IBMJ...11..215C}:
%--
\begin{equation}\label{eq:cfl}
\Delta t \lesssim \min_{i} \left( \frac{\Delta R_i}{|U_i| + c_{\mathrm{s},i}} \right),
\end{equation}
%--
where $c_{\mathrm{s},i}=\sqrt{P_i/\rho_i}$ denotes the local sound speed. Typically, the bulk velocity $U$ remains much smaller than $c_\mathrm{s}$. However, in the SMFP regime, particularly in regions of rapid expansion (as shown in Figure \ref{fig:fig4}), the bulk velocity can become comparable to the local sound speed. This raises a validity issue regarding the assumption of local thermodynamic equilibrium.

Physically, equilibrium is maintained only if the collisional relaxation timescale, $t_\mathrm{r} \sim (\rho c_\mathrm{s} \sigma)^{-1}$, is significantly shorter than the dynamical timescale of the bulk motion, characterized by the shell crossing time $t_\mathrm{sc} \sim \Delta R/U$. At radii $R\sim10^{-1}R_\mathrm{s}$, the combination of low density and high bulk velocity results in these two timescales becoming comparable ($t_\mathrm{r} \sim t_\mathrm{sc}$). In reality, if the expansion speed is too high, the collision rate becomes insufficient to maintain thermal equilibrium or effectively transport heat. Consequently, the outer shells would not gain thermal energy as efficiently as predicted by equilibrium theory, and the expansion would naturally be self-regulated and suppressed. This mechanism prevents the system from entering an unphysical regime where thermally driven expansion exceeds the sound speed. However, in our standard hydrodynamical framework, thermal equilibrium is enforcing instantaneously via the equation of state (Eq. \eqref{eq:ms9}), which lacks the intrinsic physical mechanism to dampen the expansion when the collision rate is insufficient. While a rigorous treatment requires solving the full collisional Boltzmann equation \citep[see e.g.,][]{Gurian:2025zpc}, we adopt a simplified approach here. We introduce a phenomenological pressure modification factor, $f_\mathrm{P}$, to account for this non-equilibrium effect. This factor reduces the effective pressure when the relaxation timescale becomes significant:
%--
\begin{equation}\label{eq:pfac}
f_\mathrm{P}=\left( \frac{t_\mathrm{r}}{t_\mathrm{sc}}+1\right)^{-1} .
\end{equation}
%--

\bibliography{refs}

@PREAMBLE{
 "\providecommand{\noopsort}[1]{}" 
 # "\providecommand{\singleletter}[1]{#1}%" 
}

@article{SDSS:2001emm,
    author = "Fan, Xiaohui and others",
    collaboration = "SDSS",
    title = "{A Survey of z \ensuremath{>} 5.8 quasars in the Sloan Digital Sky Survey I: Discovery of three new quasars and the spatial density of luminous quasars at z \textasciitilde{} 6}",
    eprint = "astro-ph/0108063",
    archivePrefix = "arXiv",
    doi = "10.1086/324111",
    journal = "Astron. J.",
    volume = "122",
    pages = "2833",
    year = "2001"
}

@article{Fan:2022fhc,
    author = "Fan, Xiaohui and Banados, Eduardo and Simcoe, Robert A.",
    title = "{Quasars and the Intergalactic Medium at Cosmic Dawn}",
    eprint = "2212.06907",
    archivePrefix = "arXiv",
    primaryClass = "astro-ph.GA",
    doi = "10.1146/annurev-astro-052920-102455",
    journal = "Ann. Rev. Astron. Astrophys.",
    volume = "61",
    pages = "373--426",
    year = "2023"
}

@article{Willott:2003xf,
    author = "Willott, Chris J. and McLure, Ross J. and Jarvis, Matt J.",
    title = "{A 3x10**9 solar mass black hole in the quasar SDSS J1148+5251 at z=6.41}",
    eprint = "astro-ph/0303062",
    archivePrefix = "arXiv",
    doi = "10.1086/375126",
    journal = "Astrophys. J. Lett.",
    volume = "587",
    pages = "L15--L18",
    year = "2003"
}

@article{Heger:2002by,
    author = "Heger, Alexander and Fryer, C. L. and Woosley, S. E. and Langer, N. and Hartmann, D. H.",
    title = "{How massive single stars end their life}",
    eprint = "astro-ph/0212469",
    archivePrefix = "arXiv",
    doi = "10.1086/375341",
    journal = "Astrophys. J.",
    volume = "591",
    pages = "288--300",
    year = "2003"
}

@article{Kauffmann:1999ce,
    author = "Kauffmann, Guinevere and Haehnelt, Martin",
    title = "{A Unified model for the evolution of galaxies and quasars}",
    eprint = "astro-ph/9906493",
    archivePrefix = "arXiv",
    doi = "10.1046/j.1365-8711.2000.03077.x",
    journal = "Mon. Not. Roy. Astron. Soc.",
    volume = "311",
    pages = "576--588",
    year = "2000"
}

@article{Inayoshi:2019fun,
    author = "Inayoshi, Kohei and Visbal, Eli and Haiman, Zolt\'an",
    title = "{The Assembly of the First Massive Black Holes}",
    eprint = "1911.05791",
    archivePrefix = "arXiv",
    primaryClass = "astro-ph.GA",
    doi = "10.1146/annurev-astro-120419-014455",
    journal = "Ann. Rev. Astron. Astrophys.",
    volume = "58",
    pages = "27--97",
    year = "2020"
}

@article{Gardner:2006ky,
    author = "Gardner, Jonathan P. and others",
    title = "{The James Webb Space Telescope}",
    eprint = "astro-ph/0606175",
    archivePrefix = "arXiv",
    doi = "10.1007/s11214-006-8315-7",
    journal = "Space Sci. Rev.",
    volume = "123",
    pages = "485",
    year = "2006"
}

@article{Bhowmick:2024jwc,
    author = "Bhowmick, Aklant K. and others",
    title = "{Growth of high-redshift supermassive black holes from heavy seeds in the BRAHMA cosmological simulations: implications of overmassive black holes}",
    eprint = "2406.14658",
    archivePrefix = "arXiv",
    primaryClass = "astro-ph.GA",
    doi = "10.1093/mnras/stae1819",
    journal = "Mon. Not. Roy. Astron. Soc.",
    volume = "533",
    number = "2",
    pages = "1907--1926",
    year = "2024"
}

@article{Matthee:2023utn,
    author = "Matthee, Jorryt and others",
    title = "{Little Red Dots: An Abundant Population of Faint Active Galactic Nuclei at z \ensuremath{\sim} 5 Revealed by the EIGER and FRESCO JWST Surveys}",
    eprint = "2306.05448",
    archivePrefix = "arXiv",
    primaryClass = "astro-ph.GA",
    doi = "10.3847/1538-4357/ad2345",
    journal = "Astrophys. J.",
    volume = "963",
    number = "2",
    pages = "129",
    year = "2024"
}

@article{Oppenheimer:1939ue,
    author = "Oppenheimer, J. R. and Snyder, H.",
    title = "{On Continued gravitational contraction}",
    doi = "10.1103/PhysRev.56.455",
    journal = "Phys. Rev.",
    volume = "56",
    pages = "455--459",
    year = "1939"
}

@article{Bogdan:2023ilu,
    author = "Bogdan, Akos and others",
    title = "{Evidence for heavy-seed origin of early supermassive black holes from a z\,\ensuremath{\approx}\,10 X-ray quasar}",
    eprint = "2305.15458",
    archivePrefix = "arXiv",
    primaryClass = "astro-ph.GA",
    doi = "10.1038/s41550-023-02111-9",
    journal = "Nature Astron.",
    volume = "8",
    number = "1",
    pages = "126--133",
    year = "2024"
}

@article{Goulding:2023gqa,
    author = "Goulding, Andy D. and others",
    title = "{UNCOVER: The Growth of the First Massive Black Holes from JWST/NIRSpec\textemdash{}Spectroscopic Redshift Confirmation of an X-Ray Luminous AGN at z = 10.1}",
    eprint = "2308.02750",
    archivePrefix = "arXiv",
    primaryClass = "astro-ph.GA",
    doi = "10.3847/2041-8213/acf7c5",
    journal = "Astrophys. J. Lett.",
    volume = "955",
    number = "1",
    pages = "L24",
    year = "2023"
}

@article{Volonteri:2010wz,
    author = "Volonteri, Marta",
    title = "{Formation of Supermassive Black Holes}",
    eprint = "1003.4404",
    archivePrefix = "arXiv",
    primaryClass = "astro-ph.CO",
    doi = "10.1007/s00159-010-0029-x",
    journal = "Astron. Astrophys. Rev.",
    volume = "18",
    pages = "279--315",
    year = "2010"
}

@article{Salpeter:1964kb,
    author = "Salpeter, E. E.",
    title = "{Accretion of Interstellar Matter by Massive Objects}",
    doi = "10.1086/147973",
    journal = "Astrophys. J.",
    volume = "140",
    pages = "796--800",
    year = "1964"
}

@article{Hopkins:2009td,
    author = "Hopkins, Philip F. and Quataert, Eliot",
    title = "{How Do Massive Black Holes Get Their Gas?}",
    eprint = "0912.3257",
    archivePrefix = "arXiv",
    primaryClass = "astro-ph.CO",
    doi = "10.1111/j.1365-2966.2010.17064.x",
    journal = "Mon. Not. Roy. Astron. Soc.",
    volume = "407",
    pages = "1529--1564",
    year = "2010"
}

@article{Barack:2018yly,
    author = "Barack, Leor and others",
    title = "{Black holes, gravitational waves and fundamental physics: a roadmap}",
    eprint = "1806.05195",
    archivePrefix = "arXiv",
    primaryClass = "gr-qc",
    doi = "10.1088/1361-6382/ab0587",
    journal = "Class. Quant. Grav.",
    volume = "36",
    number = "14",
    pages = "143001",
    year = "2019"
}

@article{Chantavat:2023dfg,
    author = "Chantavat, Teeraparb and Chongchitnan, Siri and Silk, Joseph",
    title = "{The most massive Population III stars}",
    eprint = "2302.09763",
    archivePrefix = "arXiv",
    primaryClass = "astro-ph.SR",
    doi = "10.1093/mnras/stad1196",
    journal = "Mon. Not. Roy. Astron. Soc.",
    volume = "522",
    number = "3",
    pages = "3256--3262",
    year = "2023"
}

@article{Begelman:2006db,
    author = "Begelman, Mitchell C. and Volonteri, Marta and Rees, Martin J.",
    title = "{Formation of supermassive black holes by direct collapse in pregalactic halos}",
    eprint = "astro-ph/0602363",
    archivePrefix = "arXiv",
    doi = "10.1111/j.1365-2966.2006.10467.x",
    journal = "Mon. Not. Roy. Astron. Soc.",
    volume = "370",
    pages = "289--298",
    year = "2006"
}

@article{Bromm:2002hb,
    author = "Bromm, Volker and Loeb, Abraham",
    title = "{Formation of the first supermassive black holes}",
    eprint = "astro-ph/0212400",
    archivePrefix = "arXiv",
    doi = "10.1086/377529",
    journal = "Astrophys. J.",
    volume = "596",
    pages = "34--46",
    year = "2003"
}

@article{Shang:2009ij,
    author = "Shang, Cien and Bryan, Greg and Haiman, Zoltan",
    title = "{Supermassive Black Hole Formation by Direct Collapse: Keeping Protogalactic Gas H\_2--Free in Dark Matter Halos with Virial Temperatures T\_vir \ensuremath{>}\textasciitilde{} 10\textasciicircum{}4 K}",
    eprint = "0906.4773",
    archivePrefix = "arXiv",
    primaryClass = "astro-ph.CO",
    doi = "10.1111/j.1365-2966.2009.15960.x",
    journal = "Mon. Not. Roy. Astron. Soc.",
    volume = "402",
    pages = "1249",
    year = "2010"
}

@article{Lodato:2006hw,
    author = "Lodato, Giuseppe and Natarajan, Priya",
    title = "{Supermassive black hole formation during the assembly of pre-galactic discs}",
    eprint = "astro-ph/0606159",
    archivePrefix = "arXiv",
    doi = "10.1111/j.1365-2966.2006.10801.x",
    journal = "Mon. Not. Roy. Astron. Soc.",
    volume = "371",
    pages = "1813--1823",
    year = "2006"
}

@article{Glover:2000ph,
    author = "Glover, Simon C. O. and Brand, Peter W. J. L.",
    title = "{On the photodissociation of H(2) by the first stars}",
    eprint = "astro-ph/0005576",
    archivePrefix = "arXiv",
    doi = "10.1046/j.1365-8711.2001.03993.x",
    journal = "Mon. Not. Roy. Astron. Soc.",
    volume = "321",
    pages = "385--397",
    year = "2001"
}

@article{Latif:2013dua,
    author = "Latif, M. A. and Schleicher, D. R. G. and Schmidt, W. and Niemeyer, J.",
    title = "{Black hole formation in the early universe}",
    eprint = "1304.0962",
    archivePrefix = "arXiv",
    primaryClass = "astro-ph.CO",
    doi = "10.1093/mnras/stt834",
    journal = "Mon. Not. Roy. Astron. Soc.",
    volume = "433",
    pages = "1607",
    year = "2013"
}

@article{Latif:2016qau,
    author = "Latif, Muhammad A. and Ferrara, Andrea",
    title = "{Formation of supermassive black hole seeds}",
    eprint = "1605.07391",
    archivePrefix = "arXiv",
    primaryClass = "astro-ph.GA",
    doi = "10.1017/pasa.2016.41",
    journal = "Publ. Astron. Soc. Austral.",
    volume = "33",
    pages = "e051",
    year = "2016"
}

@article{Kritos:2024upo,
    author = "Kritos, Konstantinos and Berti, Emanuele and Silk, Joseph",
    title = "{Supermassive black holes from runaway mergers and accretion in nuclear star clusters}",
    eprint = "2404.11676",
    archivePrefix = "arXiv",
    primaryClass = "astro-ph.HE",
    doi = "10.1093/mnras/stae1145",
    journal = "Mon. Not. Roy. Astron. Soc.",
    volume = "531",
    number = "1",
    pages = "133--136",
    year = "2024"
}

@article{Roberts:2024wup,
    author = "Roberts, M. Grant and others",
    title = "{Early formation of supermassive black holes from the collapse of strongly self-interacting dark matter}",
    eprint = "2410.17480",
    archivePrefix = "arXiv",
    primaryClass = "astro-ph.GA",
    doi = "10.1088/1475-7516/2025/01/060",
    journal = "JCAP",
    volume = "01",
    pages = "060",
    year = "2025"
}

@article{Santoro:2005tv,
    author = "Santoro, Fernando and Shull, J. Michael",
    title = "{Critical metallicity and fine-structure emission of primordial gas enriched by the first stars}",
    eprint = "astro-ph/0509101",
    archivePrefix = "arXiv",
    doi = "10.1086/501518",
    journal = "Astrophys. J.",
    volume = "643",
    pages = "26--37",
    year = "2006"
}

@article{Devecchi:2012nw,
    author = "Devecchi, B. and Volonteri, M. and Rossi, E. M. and Colpi, M. and Portegies Zwart, S.",
    title = "{High-redshift formation and evolution of central massive objects II: The census of BH seeds}",
    eprint = "1201.3761",
    archivePrefix = "arXiv",
    primaryClass = "astro-ph.CO",
    doi = "10.1111/j.1365-2966.2012.20406.x",
    journal = "Mon. Not. Roy. Astron. Soc.",
    volume = "421",
    pages = "1465",
    year = "2012"
}

@article{Inayoshi:2015pox,
    author = "Inayoshi, Kohei and Haiman, Zoltan and Ostriker, Jeremiah P.",
    title = "{Hyper-Eddington accretion flows on to massive black holes}",
    eprint = "1511.02116",
    archivePrefix = "arXiv",
    primaryClass = "astro-ph.HE",
    doi = "10.1093/mnras/stw836",
    journal = "Mon. Not. Roy. Astron. Soc.",
    volume = "459",
    number = "4",
    pages = "3738--3755",
    year = "2016"
}

@article{Shi:2022hud,
    author = "Shi, Yanlong and Kremer, Kyle and Grudi\'c, Michael Y. and Gerling-Dunsmore, Hannalore J. and Hopkins, Philip F.",
    title = "{Hyper-Eddington black hole growth in star-forming molecular clouds and galactic nuclei: can it happen?}",
    eprint = "2208.05025",
    archivePrefix = "arXiv",
    primaryClass = "astro-ph.GA",
    doi = "10.1093/mnras/stac3245",
    journal = "Mon. Not. Roy. Astron. Soc.",
    volume = "518",
    number = "3",
    pages = "3606--3621",
    year = "2022"
}

@article{Johnson:2012cw,
    author = "Johnson, Jarrett L. and Whalen, Daniel J. and Li, Hui and Holz, Daniel E.",
    title = "{Supermassive Seeds for Supermassive Black Holes}",
    eprint = "1211.0548",
    archivePrefix = "arXiv",
    primaryClass = "astro-ph.CO",
    reportNumber = "LA-UR-12-25328",
    doi = "10.1088/0004-637X/771/2/116",
    journal = "Astrophys. J.",
    volume = "771",
    pages = "116",
    year = "2013"
}

@article{Bondi:1952ni,
    author = "Bondi, H.",
    title = "{On spherically symmetrical accretion}",
    doi = "10.1093/mnras/112.2.195",
    journal = "Mon. Not. Roy. Astron. Soc.",
    volume = "112",
    pages = "195",
    year = "1952"
}

@article{Haiman:2004ve,
    author = "Haiman, Zoltan",
    title = "{Constraints from gravitational recoil on the growth of supermassive black holes at high redshift}",
    eprint = "astro-ph/0404196",
    archivePrefix = "arXiv",
    doi = "10.1086/422910",
    journal = "Astrophys. J.",
    volume = "613",
    pages = "36--40",
    year = "2004"
}

@article{Yoo:2004ze,
    author = "Yoo, Jaiyul and Miralda-Escude, Jordi",
    title = "{Formation of the black holes in the highest redshift quasars}",
    eprint = "astro-ph/0406217",
    archivePrefix = "arXiv",
    doi = "10.1086/425416",
    journal = "Astrophys. J. Lett.",
    volume = "614",
    pages = "L25--L28",
    year = "2004"
}

@article{Zeldovich:1967lct,
    author = "Zel'dovich, Ya. B. and Novikov, I. D.",
    title = "{The Hypothesis of Cores Retarded during Expansion and the Hot Cosmological Model}",
    journal = "Sov. Astron.",
    volume = "10",
    pages = "602",
    year = "1967"
}

@article{Hawking:1971ei,
    author = "Hawking, Stephen",
    title = "{Gravitationally collapsed objects of very low mass}",
    doi = "10.1093/mnras/152.1.75",
    journal = "Mon. Not. Roy. Astron. Soc.",
    volume = "152",
    pages = "75",
    year = "1971"
}

@article{Carr:1974nx,
    author = "Carr, Bernard J. and Hawking, S. W.",
    title = "{Black holes in the early Universe}",
    doi = "10.1093/mnras/168.2.399",
    journal = "Mon. Not. Roy. Astron. Soc.",
    volume = "168",
    pages = "399--415",
    year = "1974"
}

@article{Khlopov:2008qy,
    author = "Khlopov, Maxim Yu.",
    title = "{Primordial Black Holes}",
    eprint = "0801.0116",
    archivePrefix = "arXiv",
    primaryClass = "astro-ph",
    doi = "10.1088/1674-4527/10/6/001",
    journal = "Res. Astron. Astrophys.",
    volume = "10",
    pages = "495--528",
    year = "2010"
}

@article{Escriva:2022duf,
    author = "Escriv\`a, Albert and Kuhnel, Florian and Tada, Yuichiro",
    editor = "Sedda, Manuel Arca and Bortolas, Elisa and Spera, Mario",
    title = "{Primordial Black Holes}",
    eprint = "2211.05767",
    archivePrefix = "arXiv",
    primaryClass = "astro-ph.CO",
    doi = "10.1016/B978-0-32-395636-9.00012-8",
    month = "11",
    year = "2022"
}

@article{Escriva:2021aeh,
    author = "Escriv\`a, Albert",
    title = "{PBH Formation from Spherically Symmetric Hydrodynamical Perturbations: A Review}",
    eprint = "2111.12693",
    archivePrefix = "arXiv",
    primaryClass = "gr-qc",
    doi = "10.3390/universe8020066",
    journal = "Universe",
    volume = "8",
    number = "2",
    pages = "66",
    year = "2022"
}

@article{Carr:2018rid,
    author = "Carr, Bernard and Silk, Joseph",
    title = "{Primordial Black Holes as Generators of Cosmic Structures}",
    eprint = "1801.00672",
    archivePrefix = "arXiv",
    primaryClass = "astro-ph.CO",
    doi = "10.1093/mnras/sty1204",
    journal = "Mon. Not. Roy. Astron. Soc.",
    volume = "478",
    number = "3",
    pages = "3756--3775",
    year = "2018"
}

@article{Inomata:2016uip,
    author = "Inomata, Keisuke and Kawasaki, Masahiro and Tada, Yuichiro",
    title = "{Revisiting constraints on small scale perturbations from big-bang nucleosynthesis}",
    eprint = "1605.04646",
    archivePrefix = "arXiv",
    primaryClass = "astro-ph.CO",
    reportNumber = "IPMU-16-0068",
    doi = "10.1103/PhysRevD.94.043527",
    journal = "Phys. Rev. D",
    volume = "94",
    number = "4",
    pages = "043527",
    year = "2016"
}

@article{Cirelli:2024ssz,
    author = "Cirelli, Marco and Strumia, Alessandro and Zupan, Jure",
    title = "{Dark Matter}",
    eprint = "2406.01705",
    archivePrefix = "arXiv",
    primaryClass = "hep-ph",
    month = "6",
    year = "2024"
}

@article{Freeman:1970mx,
    author = "Freeman, K. C.",
    title = "{On the disks of spiral and SO Galaxies}",
    doi = "10.1086/150474",
    journal = "Astrophys. J.",
    volume = "160",
    pages = "811",
    year = "1970"
}

@article{Brainerd:1995da,
    author = "Brainerd, Tereasa G. and Blandford, Roger D. and Smail, Ian",
    title = "{Measuring galaxy masses using galaxy - galaxy gravitational lensing}",
    eprint = "astro-ph/9503073",
    archivePrefix = "arXiv",
    reportNumber = "LANL-TGB-9501",
    doi = "10.1086/177537",
    journal = "Astrophys. J.",
    volume = "466",
    pages = "623",
    year = "1996"
}

@BOOK{1980lssu.book.....P,
       author = {{Peebles}, P.~J.~E.},
        title = "{The large-scale structure of the universe}",
         year = 1980,
       adsurl = {https://ui.adsabs.harvard.edu/abs/1980lssu.book.....P}
}

@article{WMAP:2010qai,
    author = "Komatsu, E. and others",
    collaboration = "WMAP",
    title = "{Seven-Year Wilkinson Microwave Anisotropy Probe (WMAP) Observations: Cosmological Interpretation}",
    eprint = "1001.4538",
    archivePrefix = "arXiv",
    primaryClass = "astro-ph.CO",
    doi = "10.1088/0067-0049/192/2/18",
    journal = "Astrophys. J. Suppl.",
    volume = "192",
    pages = "18",
    year = "2011"
}

@article{Planck:2018vyg,
    author = "Aghanim, N. and others",
    collaboration = "Planck",
    title = "{Planck 2018 results. VI. Cosmological parameters}",
    eprint = "1807.06209",
    archivePrefix = "arXiv",
    primaryClass = "astro-ph.CO",
    doi = "10.1051/0004-6361/201833910",
    journal = "Astron. Astrophys.",
    volume = "641",
    pages = "A6",
    year = "2020",
    note = "[Erratum: Astron.Astrophys. 652, C4 (2021)]"
}

@BOOK{2010gfe..book.....M,
       author = {{Mo}, Houjun and {van den Bosch}, Frank C. and {White}, Simon},
        title = "{Galaxy Formation and Evolution}",
         year = 2010,
          doi = {10.1017/CBO9780511807244},
       adsurl = {https://ui.adsabs.harvard.edu/abs/2010gfe..book.....M}
}

@article{Davis:1985rj,
    author = "Davis, Marc and Efstathiou, George and Frenk, Carlos S. and White, Simon D. M.",
    editor = "Srednicki, M. A.",
    title = "{The Evolution of Large Scale Structure in a Universe Dominated by Cold Dark Matter}",
    reportNumber = "NSF-ITP-84-129",
    doi = "10.1086/163168",
    journal = "Astrophys. J.",
    volume = "292",
    pages = "371--394",
    year = "1985"
}

@article{Vogelsberger:2019ynw,
    author = "Vogelsberger, Mark and Marinacci, Federico and Torrey, Paul and Puchwein, Ewald",
    title = "{Cosmological Simulations of Galaxy Formation}",
    eprint = "1909.07976",
    archivePrefix = "arXiv",
    primaryClass = "astro-ph.GA",
    doi = "10.1038/s42254-019-0127-2",
    journal = "Nature Rev. Phys.",
    volume = "2",
    number = "1",
    pages = "42--66",
    year = "2020"
}

@article{Lynden-Bell:1966zjv,
    author = "Lynden-Bell, Donald",
    title = "{Statistical mechanics of violent relaxation in stellar systems}",
    journal = "Mon. Not. Roy. Astron. Soc.",
    volume = "136",
    pages = "101--121",
    year = "1967"
}

@article{Arguelles:2020qsi,
    author = {Arg\"uelles, Carlos R. and D\'\i{}az, Manuel I. and Krut, Andreas and Yunis, Rafael},
    title = "{On the formation and stability of fermionic dark matter haloes in a cosmological framework}",
    eprint = "2012.11709",
    archivePrefix = "arXiv",
    primaryClass = "astro-ph.GA",
    doi = "10.1093/mnras/staa3986",
    journal = "Mon. Not. Roy. Astron. Soc.",
    volume = "502",
    number = "3",
    pages = "4227--4246",
    year = "2021"
}

@article{Bullock:2017xww,
    author = "Bullock, James S. and Boylan-Kolchin, Michael",
    title = "{Small-Scale Challenges to the $\Lambda$CDM Paradigm}",
    eprint = "1707.04256",
    archivePrefix = "arXiv",
    primaryClass = "astro-ph.CO",
    doi = "10.1146/annurev-astro-091916-055313",
    journal = "Ann. Rev. Astron. Astrophys.",
    volume = "55",
    pages = "343--387",
    year = "2017"
}

@article{Spergel:1999mh,
    author = "Spergel, David N. and Steinhardt, Paul J.",
    title = "{Observational evidence for selfinteracting cold dark matter}",
    eprint = "astro-ph/9909386",
    archivePrefix = "arXiv",
    doi = "10.1103/PhysRevLett.84.3760",
    journal = "Phys. Rev. Lett.",
    volume = "84",
    pages = "3760--3763",
    year = "2000"
}

@article{Balberg:2002ue,
    author = "Balberg, Shmuel and Shapiro, Stuart L. and Inagaki, Shogo",
    title = "{Selfinteracting dark matter halos and the gravothermal catastrophe}",
    eprint = "astro-ph/0110561",
    archivePrefix = "arXiv",
    doi = "10.1086/339038",
    journal = "Astrophys. J.",
    volume = "568",
    pages = "475--487",
    year = "2002"
}

@article{Jiang:2025jtr,
    author = "Jiang, Fangzhou and Jia, Zixiang and Zheng, Haonan and Ho, Luis C. and Inayoshi, Kohei and Shen, Xuejian and Vogelsberger, Mark and Feng, Wei-Xiang",
    title = "{Formation of the Little Red Dots from the Core-collapse of Self-interacting Dark Matter Halos}",
    eprint = "2503.23710",
    archivePrefix = "arXiv",
    primaryClass = "astro-ph.GA",
    month = "3",
    year = "2025"
}

@article{Vogelsberger:2012ku,
    author = "Vogelsberger, Mark and Zavala, Jesus and Loeb, Abraham",
    title = "{Subhaloes in Self-Interacting Galactic Dark Matter Haloes}",
    eprint = "1201.5892",
    archivePrefix = "arXiv",
    primaryClass = "astro-ph.CO",
    doi = "10.1111/j.1365-2966.2012.21182.x",
    journal = "Mon. Not. Roy. Astron. Soc.",
    volume = "423",
    pages = "3740",
    year = "2012"
}

@article{Oman:2015xda,
    author = "Oman, Kyle A. and others",
    title = "{The unexpected diversity of dwarf galaxy rotation curves}",
    eprint = "1504.01437",
    archivePrefix = "arXiv",
    primaryClass = "astro-ph.GA",
    doi = "10.1093/mnras/stv1504",
    journal = "Mon. Not. Roy. Astron. Soc.",
    volume = "452",
    number = "4",
    pages = "3650--3665",
    year = "2015"
}

@article{Zavala:2012us,
    author = "Zavala, Jesus and Vogelsberger, Mark and Walker, Matthew G.",
    title = "{Constraining Self-Interacting Dark Matter with the Milky Way's dwarf spheroidals}",
    eprint = "1211.6426",
    archivePrefix = "arXiv",
    primaryClass = "astro-ph.CO",
    doi = "10.1093/mnrasl/sls053",
    journal = "Mon. Not. Roy. Astron. Soc.",
    volume = "431",
    pages = "L20--L24",
    year = "2013"
}

@article{Rocha:2012jg,
    author = "Rocha, Miguel and Peter, Annika H. G. and Bullock, James S. and Kaplinghat, Manoj and Garrison-Kimmel, Shea and Onorbe, Jose and Moustakas, Leonidas A.",
    title = "{Cosmological Simulations with Self-Interacting Dark Matter I: Constant Density Cores and Substructure}",
    eprint = "1208.3025",
    archivePrefix = "arXiv",
    primaryClass = "astro-ph.CO",
    doi = "10.1093/mnras/sts514",
    journal = "Mon. Not. Roy. Astron. Soc.",
    volume = "430",
    pages = "81--104",
    year = "2013"
}

@article{Peter:2012jh,
    author = "Peter, Annika H. G. and Rocha, Miguel and Bullock, James S. and Kaplinghat, Manoj",
    title = "{Cosmological Simulations with Self-Interacting Dark Matter II: Halo Shapes vs. Observations}",
    eprint = "1208.3026",
    archivePrefix = "arXiv",
    primaryClass = "astro-ph.CO",
    reportNumber = "NSF-KITP-12-147",
    doi = "10.1093/mnras/sts535",
    journal = "Mon. Not. Roy. Astron. Soc.",
    volume = "430",
    pages = "105",
    year = "2013"
}

@article{Tulin:2017ara,
    author = "Tulin, Sean and Yu, Hai-Bo",
    title = "{Dark Matter Self-interactions and Small Scale Structure}",
    eprint = "1705.02358",
    archivePrefix = "arXiv",
    primaryClass = "hep-ph",
    doi = "10.1016/j.physrep.2017.11.004",
    journal = "Phys. Rept.",
    volume = "730",
    pages = "1--57",
    year = "2018"
}

@article{Adhikari:2022sbh,
    author = "Adhikari, Susmita and others",
    title = "{Astrophysical Tests of Dark Matter Self-Interactions}",
    eprint = "2207.10638",
    archivePrefix = "arXiv",
    primaryClass = "astro-ph.CO",
    month = "7",
    year = "2022"
}

@article{Lightman:1978zz,
    author = "Lightman, Alan P. and Shapiro, Stuart L.",
    title = "{The dynamical evolution of globular clusters}",
    doi = "10.1103/RevModPhys.50.437",
    journal = "Rev. Mod. Phys.",
    volume = "50",
    pages = "437--481",
    year = "1978"
}

@article{Lynden-Bell:1980xip,
    author = "Lynden-Bell, D. and Eggleton, P. P.",
    title = "{On the consequences of the gravothermal catastrophe}",
    doi = "10.1093/mnras/191.3.483",
    journal = "Mon. Not. Roy. Astron. Soc.",
    volume = "191",
    number = "3",
    pages = "483--498",
    year = "1980"
}

@article{Pollack:2014rja,
    author = "Pollack, Jason and Spergel, David N. and Steinhardt, Paul J.",
    title = "{Supermassive Black Holes from Ultra-Strongly Self-Interacting Dark Matter}",
    eprint = "1501.00017",
    archivePrefix = "arXiv",
    primaryClass = "astro-ph.CO",
    reportNumber = "CALT-TH-2014-144",
    doi = "10.1088/0004-637X/804/2/131",
    journal = "Astrophys. J.",
    volume = "804",
    number = "2",
    pages = "131",
    year = "2015"
}

@article{Feng:2020kxv,
    author = "Feng, Wei-Xiang and Yu, Hai-Bo and Zhong, Yi-Ming",
    title = "{Seeding Supermassive Black Holes with Self-interacting Dark Matter: A Unified Scenario with Baryons}",
    eprint = "2010.15132",
    archivePrefix = "arXiv",
    primaryClass = "astro-ph.CO",
    doi = "10.3847/2041-8213/ac04b0",
    journal = "Astrophys. J. Lett.",
    volume = "914",
    number = "2",
    pages = "L26",
    year = "2021"
}

@article{Xiao:2021ftk,
    author = "Xiao, Huangyu and Shen, Xuejian and Hopkins, Philip F. and Zurek, Kathryn M.",
    title = "{SMBH seeds from dissipative dark matter}",
    eprint = "2103.13407",
    archivePrefix = "arXiv",
    primaryClass = "astro-ph.CO",
    doi = "10.1088/1475-7516/2021/07/039",
    journal = "JCAP",
    volume = "07",
    pages = "039",
    year = "2021"
}

@article{Choquette:2018lvq,
    author = "Choquette, Jeremie and Cline, James M. and Cornell, Jonathan M.",
    title = "{Early formation of supermassive black holes via dark matter self-interactions}",
    eprint = "1812.05088",
    archivePrefix = "arXiv",
    primaryClass = "astro-ph.CO",
    doi = "10.1088/1475-7516/2019/07/036",
    journal = "JCAP",
    volume = "07",
    pages = "036",
    year = "2019"
}

@article{Koda:2011yb,
    author = "Koda, Jun and Shapiro, Paul R.",
    title = "{Gravothermal collapse of isolated self-interacting dark matter haloes: N-body simulation versus the fluid model}",
    eprint = "1101.3097",
    archivePrefix = "arXiv",
    primaryClass = "astro-ph.CO",
    reportNumber = "TCC-001-11",
    doi = "10.1111/j.1365-2966.2011.18684.x",
    journal = "Mon. Not. Roy. Astron. Soc.",
    volume = "415",
    pages = "1125",
    year = "2011"
}

@article{Palubski:2024ibb,
    author = "Palubski, Igor and Slone, Oren and Kaplinghat, Manoj and Lisanti, Mariangela and Jiang, Fangzhou",
    title = "{Numerical challenges in modeling gravothermal collapse in Self-Interacting Dark Matter halos}",
    eprint = "2402.12452",
    archivePrefix = "arXiv",
    primaryClass = "astro-ph.CO",
    doi = "10.1088/1475-7516/2024/09/074",
    journal = "JCAP",
    volume = "09",
    pages = "074",
    year = "2024"
}

@article{Outmezguine:2022bhq,
    author = "Outmezguine, Nadav Joseph and Boddy, Kimberly K. and Gad-Nasr, Sophia and Kaplinghat, Manoj and Sagunski, Laura",
    title = "{Universal gravothermal evolution of isolated self-interacting dark matter halos for velocity-dependent cross-sections}",
    eprint = "2204.06568",
    archivePrefix = "arXiv",
    primaryClass = "astro-ph.GA",
    doi = "10.1093/mnras/stad1705",
    journal = "Mon. Not. Roy. Astron. Soc.",
    volume = "523",
    number = "3",
    pages = "4786--4800",
    year = "2023"
}

@article{Shapiro:2014oha,
    author = "Shapiro, Stuart L. and Paschalidis, Vasileios",
    title = "{Self-interacting dark matter cusps around massive black holes}",
    eprint = "1402.0005",
    archivePrefix = "arXiv",
    primaryClass = "astro-ph.CO",
    doi = "10.1103/PhysRevD.89.023506",
    journal = "Phys. Rev. D",
    volume = "89",
    number = "2",
    pages = "023506",
    year = "2014"
}

@article{Shapiro:2018vju,
    author = "Shapiro, Stuart L.",
    title = "{Star clusters, self-interacting dark matter halos, and black hole cusps: The fluid conduction model and its extension to general relativity}",
    eprint = "1809.02618",
    archivePrefix = "arXiv",
    primaryClass = "astro-ph.HE",
    doi = "10.1103/PhysRevD.98.023021",
    journal = "Phys. Rev. D",
    volume = "98",
    number = "2",
    pages = "023021",
    year = "2018"
}

@article{Feng:2021rst,
    author = "Feng, Wei-Xiang and Yu, Hai-Bo and Zhong, Yi-Ming",
    title = "{Dynamical instability of collapsed dark matter halos}",
    eprint = "2108.11967",
    archivePrefix = "arXiv",
    primaryClass = "astro-ph.CO",
    doi = "10.1088/1475-7516/2022/05/036",
    journal = "JCAP",
    volume = "05",
    number = "05",
    pages = "036",
    year = "2022"
}

@article{Misner:1964je,
    author = "Misner, Charles W. and Sharp, David H.",
    title = "{Relativistic equations for adiabatic, spherically symmetric gravitational collapse}",
    doi = "10.1103/PhysRev.136.B571",
    journal = "Phys. Rev.",
    volume = "136",
    pages = "B571--B576",
    year = "1964"
}

@article{Hernandez:1966zia,
    author = "Hernandez, Walter C. and Misner, Charles W.",
    title = "{Observer Time as a Coordinate in Relativistic Spherical Hydrodynamics}",
    doi = "10.1086/148525",
    journal = "Astrophys. J.",
    volume = "143",
    pages = "452",
    year = "1966"
}

@article{Penrose:1964wq,
    author = "Penrose, Roger",
    title = "{Gravitational collapse and space-time singularities}",
    doi = "10.1103/PhysRevLett.14.57",
    journal = "Phys. Rev. Lett.",
    volume = "14",
    pages = "57--59",
    year = "1965"
}

@article{Hawking:1971vc,
    author = "Hawking, S. W.",
    title = "{Black holes in general relativity}",
    doi = "10.1007/BF01877517",
    journal = "Commun. Math. Phys.",
    volume = "25",
    pages = "152--166",
    year = "1972"
}

@article{Gad-Nasr:2023gvf,
    author = "Gad-Nasr, Sophia and Boddy, Kimberly K. and Kaplinghat, Manoj and Outmezguine, Nadav Joseph and Sagunski, Laura",
    title = "{On the late-time evolution of velocity-dependent self-interacting dark matter halos}",
    eprint = "2312.09296",
    archivePrefix = "arXiv",
    primaryClass = "astro-ph.GA",
    doi = "10.1088/1475-7516/2024/05/131",
    journal = "JCAP",
    volume = "05",
    pages = "131",
    year = "2024"
}

@article{Yang:2022zkd,
    author = "Yang, Shengqi and Du, Xiaolong and Zeng, Zhichao Carton and Benson, Andrew and Jiang, Fangzhou and Nadler, Ethan O. and Peter, Annika H. G.",
    title = "{Gravothermal Solutions of SIDM Halos: Mapping from Constant to Velocity-dependent Cross Section}",
    eprint = "2205.02957",
    archivePrefix = "arXiv",
    primaryClass = "astro-ph.CO",
    doi = "10.3847/1538-4357/acbd49",
    journal = "Astrophys. J.",
    volume = "946",
    number = "1",
    pages = "47",
    year = "2023"
}

@article{Correa:2020qam,
    author = "Correa, Camila A.",
    title = "{Constraining velocity-dependent self-interacting dark matter with the Milky Way\textquoteright{}s dwarf spheroidal galaxies}",
    eprint = "2007.02958",
    archivePrefix = "arXiv",
    primaryClass = "astro-ph.GA",
    doi = "10.1093/mnras/stab506",
    journal = "Mon. Not. Roy. Astron. Soc.",
    volume = "503",
    number = "1",
    pages = "920--937",
    year = "2021"
}

@article{Greene:2024phl,
    author = "Greene, Jenny E. and others",
    title = "{UNCOVER Spectroscopy Confirms the Surprising Ubiquity of Active Galactic Nuclei in Red Sources at z \ensuremath{>} 5}",
    doi = "10.3847/1538-4357/ad1e5f",
    journal = "Astrophys. J.",
    volume = "964",
    number = "1",
    pages = "39",
    year = "2024"
}

@ARTICLE{2024ApJ...968...38K,
       author = {{Kokorev}, Vasily and others},
        title = "{A Census of Photometrically Selected Little Red Dots at 4 < z < 9 in JWST Blank Fields}",
      journal = {\apj},
         year = 2024,
        month = jun,
       volume = {968},
       number = {1},
          eid = {38},
        pages = {38},
          doi = {10.3847/1538-4357/ad4265},
archivePrefix = {arXiv},
       eprint = {2401.09981},
 primaryClass = {astro-ph.GA},
       adsurl = {https://ui.adsabs.harvard.edu/abs/2024ApJ...968...38K}
}

@ARTICLE{2024arXiv240403576K,
       author = {{Kocevski}, Dale D. and others},
        title = "{The Rise of Faint, Red AGN at $z>4$: A Sample of Little Red Dots in the JWST Extragalactic Legacy Fields}",
      journal = {arXiv e-prints},
         year = 2024,
        month = apr,
          eid = {arXiv:2404.03576},
        pages = {arXiv:2404.03576},
          doi = {10.48550/arXiv.2404.03576},
archivePrefix = {arXiv},
       eprint = {2404.03576},
 primaryClass = {astro-ph.GA},
       adsurl = {https://ui.adsabs.harvard.edu/abs/2024arXiv240403576K}
}

@article{Navarro:1995iw,
    author = "Navarro, Julio F. and Frenk, Carlos S. and White, Simon D. M.",
    title = "{The Structure of cold dark matter halos}",
    eprint = "astro-ph/9508025",
    archivePrefix = "arXiv",
    doi = "10.1086/177173",
    journal = "Astrophys. J.",
    volume = "462",
    pages = "563--575",
    year = "1996"
}

@article{Navarro:1996gj,
    author = "Navarro, Julio F. and Frenk, Carlos S. and White, Simon D. M.",
    title = "{A Universal density profile from hierarchical clustering}",
    eprint = "astro-ph/9611107",
    archivePrefix = "arXiv",
    doi = "10.1086/304888",
    journal = "Astrophys. J.",
    volume = "490",
    pages = "493--508",
    year = "1997"
}

@article{Eckart:1940te,
    author = "Eckart, Carl",
    title = "{The Thermodynamics of irreversible processes. 3.. Relativistic theory of the simple fluid}",
    doi = "10.1103/PhysRev.58.919",
    journal = "Phys. Rev.",
    volume = "58",
    pages = "919--924",
    year = "1940"
}

@book{Misner:1973prb,
    author = "Misner, Charles W. and Thorne, K. S. and Wheeler, J. A.",
    title = "{Gravitation}",
    isbn = "978-0-7167-0344-0, 978-0-691-17779-3",
    publisher = "W. H. Freeman",
    address = "San Francisco",
    year = "1973"
}

@article{Tolman:1939jz,
    author = "Tolman, Richard C.",
    title = "{Static solutions of Einstein's field equations for spheres of fluid}",
    doi = "10.1103/PhysRev.55.364",
    journal = "Phys. Rev.",
    volume = "55",
    pages = "364--373",
    year = "1939"
}

@article{Oppenheimer:1939ne,
    author = "Oppenheimer, J. R. and Volkoff, G. M.",
    title = "{On massive neutron cores}",
    doi = "10.1103/PhysRev.55.374",
    journal = "Phys. Rev.",
    volume = "55",
    pages = "374--381",
    year = "1939"
}

@ARTICLE{1995ApJ...443..717B,
       author = {{Baumgarte}, Thomas W. and {Shapiro}, Stuart L. and {Teukolsky}, Saul A.},
        title = "{Computing Supernova Collapse to Neutron Stars and Black Holes}",
      journal = {\apj},
         year = 1995,
        month = apr,
       volume = {443},
        pages = {717},
          doi = {10.1086/175563},
       adsurl = {https://ui.adsabs.harvard.edu/abs/1995ApJ...443..717B}
}

@ARTICLE{1978ApJ...221..304V,
       author = {{van Riper}, K.~A.},
        title = "{The hydrodynamics of stellar collapse.}",
      journal = {\apj},
         year = 1978,
        month = apr,
       volume = {221},
        pages = {304-319},
          doi = {10.1086/156029},
       adsurl = {https://ui.adsabs.harvard.edu/abs/1978ApJ...221..304V}
}

@ARTICLE{1979ApJ...232..558V,
       author = {{van Riper}, K.~A.},
        title = "{General relativistic hydrodynamics and the adiabatic collapse of stellar cores.}",
      journal = {\apj},
         year = 1979,
        month = sep,
       volume = {232},
        pages = {558-571},
          doi = {10.1086/157314},
       adsurl = {https://ui.adsabs.harvard.edu/abs/1979ApJ...232..558V}
}

@article{May:1966zz,
    author = "May, Michael M. and White, Richard H.",
    title = "{Hydrodynamic Calculations of General-Relativistic Collapse}",
    doi = "10.1103/PhysRev.141.1232",
    journal = "Phys. Rev.",
    volume = "141",
    pages = "1232--1241",
    year = "1966"
}

@book{Hawking:1973uf,
    author = "Hawking, Stephen W. and Ellis, George F. R.",
    title = "{The Large Scale Structure of Space-Time}",
    doi = "10.1017/9781009253161",
    isbn = "978-1-009-25316-1, 978-1-009-25315-4, 978-0-521-20016-5, 978-0-521-09906-6, 978-0-511-82630-6, 978-0-521-09906-6",
    publisher = "Cambridge University Press",
    series = "Cambridge Monographs on Mathematical Physics",
    month = "2",
    year = "2023"
}

@article{Liebendoerfer:2000fw,
    author = "Liebendoerfer, Matthias and Mezzacappa, Anthony and Thielemann, Friedrich-Karl",
    title = "{Conservative general relativistic radiation hydrodynamics in spherical symmetry and comoving coordinates}",
    eprint = "astro-ph/0012201",
    archivePrefix = "arXiv",
    doi = "10.1103/PhysRevD.63.104003",
    journal = "Phys. Rev. D",
    volume = "63",
    pages = "104003",
    year = "2001"
}

@article{Balberg:2001qg,
    author = "Balberg, Shmuel and Shapiro, Stuart L.",
    title = "{Gravothermal collapse of selfinteracting dark matter halos and the origin of massive black holes}",
    eprint = "astro-ph/0111176",
    archivePrefix = "arXiv",
    doi = "10.1103/PhysRevLett.88.101301",
    journal = "Phys. Rev. Lett.",
    volume = "88",
    pages = "101301",
    year = "2002"
}

@article{Ahn:2004xt,
    author = "Ahn, Kyung-Jin and Shapiro, Paul R.",
    title = "{Formation and evolution of the self-interacting dark matter halos}",
    eprint = "astro-ph/0412169",
    archivePrefix = "arXiv",
    doi = "10.1111/j.1365-2966.2005.09492.x",
    journal = "Mon. Not. Roy. Astron. Soc.",
    volume = "363",
    pages = "1092--1124",
    year = "2005"
}

@article{DiCintio:2017zdz,
    author = "Di Cintio, Arianna and Tremmel, Michael and Governato, Fabio and Pontzen, Andrew and Zavala, Jes{\'u}s and Bastidas Fry, Alexander and Brooks, Alyson and Vogelsberger, Mark",
    title = "{A rumble in the dark: signatures of self-interacting dark matter in supermassive black hole dynamics and galaxy density profiles}",
    eprint = "1701.04410",
    archivePrefix = "arXiv",
    primaryClass = "astro-ph.GA",
    doi = "10.1093/mnras/stx1043",
    journal = "Mon. Not. Roy. Astron. Soc.",
    volume = "469",
    number = "3",
    pages = "2845--2854",
    year = "2017"
}

@article{Banerjee:2019bjp,
    author = "Banerjee, Arka and Adhikari, Susmita and Dalal, Neal and More, Surhud and Kravtsov, Andrey",
    title = "{Signatures of Self-Interacting dark matter on cluster density profile and subhalo distributions}",
    eprint = "1906.12026",
    archivePrefix = "arXiv",
    primaryClass = "astro-ph.CO",
    doi = "10.1088/1475-7516/2020/02/024",
    journal = "JCAP",
    volume = "02",
    pages = "024",
    year = "2020"
}

@article{Kaplinghat:2019dhn,
    author = "Kaplinghat, Manoj and Ren, Tao and Yu, Hai-Bo",
    title = "{Dark Matter Cores and Cusps in Spiral Galaxies and their Explanations}",
    eprint = "1911.00544",
    archivePrefix = "arXiv",
    primaryClass = "astro-ph.GA",
    doi = "10.1088/1475-7516/2020/06/027",
    journal = "JCAP",
    volume = "06",
    pages = "027",
    year = "2020"
}

@article{Yang:2021kdf,
    author = "Yang, Daneng and Yu, Hai-Bo",
    title = "{Self-interacting dark matter and small-scale gravitational lenses in galaxy clusters}",
    eprint = "2102.02375",
    archivePrefix = "arXiv",
    primaryClass = "astro-ph.GA",
    doi = "10.1103/PhysRevD.104.103031",
    journal = "Phys. Rev. D",
    volume = "104",
    number = "10",
    pages = "103031",
    year = "2021"
}

@article{Shen:2021frv,
    author = "Shen, Xuejian and Hopkins, Philip F. and Necib, Lina and Jiang, Fangzhou and Boylan-Kolchin, Michael and Wetzel, Andrew",
    title = "{Dissipative dark matter on FIRE {\textendash} I. Structural and kinematic properties of dwarf galaxies}",
    eprint = "2102.09580",
    archivePrefix = "arXiv",
    primaryClass = "astro-ph.GA",
    doi = "10.1093/mnras/stab2042",
    journal = "Mon. Not. Roy. Astron. Soc.",
    volume = "506",
    number = "3",
    pages = "4421--4445",
    year = "2021"
}

@article{Gilman:2021sdr,
    author = "Gilman, Daniel and Bovy, Jo and Treu, Tommaso and Nierenberg, Anna and Birrer, Simon and Benson, Andrew and Sameie, Omid",
    title = "{Strong lensing signatures of self-interacting dark matter in low-mass haloes}",
    eprint = "2105.05259",
    archivePrefix = "arXiv",
    primaryClass = "astro-ph.CO",
    doi = "10.1093/mnras/stab2335",
    journal = "Mon. Not. Roy. Astron. Soc.",
    volume = "507",
    number = "2",
    pages = "2432--2447",
    year = "2021"
}

@article{Meshveliani:2022rih,
    author = "Meshveliani, Tamar and Zavala, Jes{\'u}s and Lovell, Mark R.",
    title = "{Gravothermal collapse of self-interacting dark matter halos as the origin of intermediate mass black holes in Milky~Way satellites}",
    eprint = "2210.01817",
    archivePrefix = "arXiv",
    primaryClass = "astro-ph.GA",
    doi = "10.1103/PhysRevD.107.083010",
    journal = "Phys. Rev. D",
    volume = "107",
    number = "8",
    pages = "083010",
    year = "2023"
}

@article{Arguelles:2023hab,
    author = {Arg{\"u}elles, C. R. and Boshkayev, K. and Krut, A. and Nurbakhyt, G. and Rueda, J. A. and Ruffini, R. and Uribe-Su{\'a}rez, J. D. and Yunis, R.},
    title = "{On the growth of supermassive black holes formed from the gravitational collapse of fermionic dark matter cores}",
    eprint = "2305.02430",
    archivePrefix = "arXiv",
    primaryClass = "astro-ph.CO",
    doi = "10.1093/mnras/stad1380",
    journal = "Mon. Not. Roy. Astron. Soc.",
    volume = "523",
    number = "2",
    pages = "2209--2218",
    year = "2023"
}

@article{Zhong:2023yzk,
    author = "Zhong, Yi-Ming and Yang, Daneng and Yu, Hai-Bo",
    title = "{The impact of baryonic potentials on the gravothermal evolution of self-interacting dark matter haloes}",
    eprint = "2306.08028",
    archivePrefix = "arXiv",
    primaryClass = "astro-ph.CO",
    doi = "10.1093/mnras/stad2765",
    journal = "Mon. Not. Roy. Astron. Soc.",
    volume = "526",
    number = "1",
    pages = "758--770",
    year = "2023"
}

@article{Fischer:2024eaz,
    author = "Fischer, Moritz S. and Dolag, Klaus and Yu, Hai-Bo",
    title = "{Numerical challenges for energy conservation in N-body simulations of collapsing self-interacting dark matter halos}",
    eprint = "2403.00739",
    archivePrefix = "arXiv",
    primaryClass = "astro-ph.CO",
    doi = "10.1051/0004-6361/202449849",
    journal = "Astron. Astrophys.",
    volume = "689",
    pages = "A300",
    year = "2024"
}

@article{deBlok:2009sp,
    author = "de Blok, W. J. G.",
    title = "{The Core-Cusp Problem}",
    eprint = "0910.3538",
    archivePrefix = "arXiv",
    primaryClass = "astro-ph.CO",
    doi = "10.1155/2010/789293",
    journal = "Adv. Astron.",
    volume = "2010",
    pages = "789293",
    year = "2010"
}

@article{Shen:2025evo,
    author = "Shen, Tingwei and Shen, Xuejian and Xiao, Huangyu and Vogelsberger, Mark and Jiang, Fangzhou",
    title = "{Massive Black Holes Seeded by Dark Matter -- Implications for Little Red Dots and Gravitational Wave Signatures}",
    eprint = "2504.00075",
    archivePrefix = "arXiv",
    primaryClass = "astro-ph.GA",
    reportNumber = "FERMILAB-PUB-25-0224-T",
    month = "3",
    year = "2025"
}

@article{Gurian:2025zpc,
    author = "Gurian, James and May, Simon",
    title = "{Core Collapse Beyond the Fluid Approximation: The Late Evolution of Self-Interacting Dark Matter Halos}",
    eprint = "2505.15903",
    archivePrefix = "arXiv",
    primaryClass = "astro-ph.CO",
    doi = "10.1103/2ycz-3fvv",
    journal = "Phys. Rev. Lett.",
    volume = "135",
    number = "22",
    pages = "221001",
    year = "2025"
}

@article{Sabarish:2025hwb,
    author = {Sabarish, V. M. and Br{\"u}ggen, Marcus and Schmidt-Hoberg, Kai and Fischer, Moritz S.},
    title = "{Accretion of self-interacting dark matter onto supermassive black holes}",
    eprint = "2505.14779",
    archivePrefix = "arXiv",
    primaryClass = "astro-ph.CO",
    doi = "10.1051/0004-6361/202555586",
    journal = "Astron. Astrophys.",
    volume = "703",
    pages = "A142",
    year = "2025"
}

@article{Nishikawa:2019lsc,
    author = "Nishikawa, Hiroya and Boddy, Kimberly K. and Kaplinghat, Manoj",
    title = "{Accelerated core collapse in tidally stripped self-interacting dark matter halos}",
    eprint = "1901.00499",
    archivePrefix = "arXiv",
    primaryClass = "astro-ph.GA",
    doi = "10.1103/PhysRevD.101.063009",
    journal = "Phys. Rev. D",
    volume = "101",
    number = "6",
    pages = "063009",
    year = "2020"
}

@article{Essig:2018pzq,
    author = "Essig, Rouven and Mcdermott, Samuel D. and Yu, Hai-Bo and Zhong, Yi-Ming",
    title = "{Constraining Dissipative Dark Matter Self-Interactions}",
    eprint = "1809.01144",
    archivePrefix = "arXiv",
    primaryClass = "hep-ph",
    reportNumber = "FERMILAB-PUB-18-437-A",
    doi = "10.1103/PhysRevLett.123.121102",
    journal = "Phys. Rev. Lett.",
    volume = "123",
    number = "12",
    pages = "121102",
    year = "2019"
}

@article{Sameie:2018chj,
    author = "Sameie, Omid and Creasey, Peter and Yu, Hai-Bo and Sales, Laura V. and Vogelsberger, Mark and Zavala, Jesus",
    title = "{The impact of baryonic discs on the shapes and profiles of self-interacting dark matter haloes}",
    eprint = "1801.09682",
    archivePrefix = "arXiv",
    primaryClass = "astro-ph.GA",
    doi = "10.1093/mnras/sty1516",
    journal = "Mon. Not. Roy. Astron. Soc.",
    volume = "479",
    number = "1",
    pages = "359--367",
    year = "2018"
}

@article{Robertson:2018anx,
    author = "Robertson, Andrew and Harvey, David and Massey, Richard and Eke, Vincent and McCarthy, Ian G. and Jauzac, Mathilde and Li, Baojiu and Schaye, Joop",
    title = "{Observable tests of self-interacting dark matter in galaxy clusters: cosmological simulations with SIDM and baryons}",
    eprint = "1810.05649",
    archivePrefix = "arXiv",
    primaryClass = "astro-ph.CO",
    doi = "10.1093/mnras/stz1815",
    journal = "Mon. Not. Roy. Astron. Soc.",
    volume = "488",
    number = "3",
    pages = "3646--3662",
    year = "2019"
}

@BOOK{1970mtnu.book.....C,
       author = {{Chapman}, Sydeny and {Cowling}, T.~G.},
        title = "{The mathematical theory of non-uniform gases. an account of the kinetic theory of viscosity, thermal conduction and diffusion in gases}",
         year = 1970,
       adsurl = {https://ui.adsabs.harvard.edu/abs/1970mtnu.book.....C}
}

@ARTICLE{1967IBMJ...11..215C,
       author = {{Courant}, R. and {Friedrichs}, K. and {Lewy}, H.},
        title = "{On the Partial Difference Equations of Mathematical Physics}",
      journal = {IBM Journal of Research and Development},
         year = 1967,
        month = mar,
       volume = {11},
        pages = {215-234},
          doi = {10.1147/rd.112.0215},
       adsurl = {https://ui.adsabs.harvard.edu/abs/1967IBMJ...11..215C}
}

@misc{bosch2025dynamicscoresselfinteractingdark,
      title={Dynamics in the Cores of Self-Interacting Dark Matter Halos: Reduced Stalling and Accelerated Core Collapse}, 
      author={Frank C. van den Bosch and Shashank Dattathri},
      year={2025},
      eprint={2511.14912},
      archivePrefix={arXiv},
      primaryClass={astro-ph.GA},
      url={https://arxiv.org/abs/2511.14912}, 
}

@article{Feng:2025rzf,
    author = "Feng, Wei-Xiang and Yu, Hai-Bo and Zhong, Yi-Ming",
    title = "{Dark Bondi Accretion Aided by Baryons and the Origin of JWST Little Red Dots}",
    eprint = "2506.17641",
    archivePrefix = "arXiv",
    primaryClass = "astro-ph.GA",
    month = "6",
    year = "2025"
}

@article{Inayoshi:2025isg,
    author = "Inayoshi, Kohei",
    title = "{Little Red Dots as the Very First Activity of Black Hole Growth}",
    doi = "10.3847/2041-8213/adea66",
    journal = "Astrophys. J. Lett.",
    volume = "988",
    number = "1",
    pages = "L22",
    year = "2025"
}

@ARTICLE{2025ApJ...978...92L,
       author = {{Labbe}, Ivo and others},
        title = "{UNCOVER: Candidate Red Active Galactic Nuclei at 3 < z < 7 with JWST and ALMA}",
      journal = {\apj},
         year = 2025,
        month = jan,
       volume = {978},
       number = {1},
          eid = {92},
        pages = {92},
          doi = {10.3847/1538-4357/ad3551},
archivePrefix = {arXiv},
       eprint = {2306.07320},
 primaryClass = {astro-ph.GA},
       adsurl = {https://ui.adsabs.harvard.edu/abs/2025ApJ...978...92L}
}

@article{Chen:2025mzw,
    author = "Chen, Chang-Hao and Ho, Luis C. and Li, Ruancun and Zhuang, Ming-Yang",
    title = "{The Host Galaxy (If Any) of the Little Red Dots}",
    doi = "10.3847/1538-4357/ada93a",
    journal = "Astrophys. J.",
    volume = "983",
    number = "1",
    pages = "60",
    year = "2025"
}

@ARTICLE{2024A&A...691A.145M,
       author = {{Maiolino}, Roberto and others},
        title = "{JADES: The diverse population of infant black holes at 4 < z < 11: Merging, tiny, poor, but mighty}",
      journal = {\aap},
         year = 2024,
        month = nov,
       volume = {691},
          eid = {A145},
        pages = {A145},
          doi = {10.1051/0004-6361/202347640},
archivePrefix = {arXiv},
       eprint = {2308.01230},
 primaryClass = {astro-ph.GA},
       adsurl = {https://ui.adsabs.harvard.edu/abs/2024A&A...691A.145M}
}

@ARTICLE{2022NatAs...6..897S,
       author = {{Sales}, Laura V. and {Wetzel}, Andrew and {Fattahi}, Azadeh},
        title = "{Baryonic solutions and challenges for cosmological models of dwarf galaxies}",
      journal = {Nature Astronomy},
         year = 2022,
        month = jun,
       volume = {6},
        pages = {897-910},
          doi = {10.1038/s41550-022-01689-w},
archivePrefix = {arXiv},
       eprint = {2206.05295},
 primaryClass = {astro-ph.GA},
       adsurl = {https://ui.adsabs.harvard.edu/abs/2022NatAs...6..897S}
}

@ARTICLE{2006MNRAS.365..345H,
       author = {{Hu}, Jian and {Shen}, Yue and {Lou}, Yu-Qing and {Zhang}, Shuangnan},
        title = "{Forming supermassive black holes by accreting dark and baryon matter}",
      journal = {\mnras},
         year = 2006,
        month = jan,
       volume = {365},
       number = {1},
        pages = {345-351},
          doi = {10.1111/j.1365-2966.2005.09712.x},
archivePrefix = {arXiv},
       eprint = {astro-ph/0510222},
 primaryClass = {astro-ph},
       adsurl = {https://ui.adsabs.harvard.edu/abs/2006MNRAS.365..345H}
}

@ARTICLE{2026Natur.649..574R,
       author = {{Rusakov}, V. and others},
        title = "{Little red dots as young supermassive black holes in dense ionized cocoons}",
      journal = {\nat},
         year = 2026,
        month = jan,
       volume = {649},
       number = {8097},
        pages = {574-579},
          doi = {10.1038/s41586-025-09900-4},
archivePrefix = {arXiv},
       eprint = {2503.16595},
 primaryClass = {astro-ph.GA},
       adsurl = {https://ui.adsabs.harvard.edu/abs/2026Natur.649..574R}
}

@ARTICLE{2026arXiv260106024C,
       author = {{Carr}, Bernard and {Iovino}, Antonio J. and {Perna}, Gabriele and {Vaskonen}, Ville and {Veerm{\"a}e}, Hardi},
        title = "{Primordial black holes: constraints, potential evidence and prospects}",
      journal = {arXiv e-prints},
         year = 2026,
        month = jan,
          eid = {arXiv:2601.06024},
        pages = {arXiv:2601.06024},
          doi = {10.48550/arXiv.2601.06024},
archivePrefix = {arXiv},
       eprint = {2601.06024},
 primaryClass = {astro-ph.CO},
       adsurl = {https://ui.adsabs.harvard.edu/abs/2026arXiv260106024C}
}

@article{Ralegankar:2024zjd,
    author = "Ralegankar, Pranjal and Perri, Daniele and Kobayashi, Takeshi",
    title = "{Gravothermalizing into primordial black holes, boson stars, and cannibal stars}",
    eprint = "2410.18948",
    archivePrefix = "arXiv",
    primaryClass = "astro-ph.CO",
    doi = "10.1103/xpwl-w5zk",
    journal = "Phys. Rev. D",
    volume = "112",
    number = "8",
    pages = "083019",
    year = "2025"
}

@article{Arguelles:2023kqw,
    author = "Arguelles, C. R. and Rueda, J. A. and Ruffini, R.",
    title = "{Baryon-induced Collapse of Dark Matter Cores into Supermassive Black Holes}",
    eprint = "2312.07461",
    archivePrefix = "arXiv",
    primaryClass = "astro-ph.GA",
    doi = "10.3847/2041-8213/ad1490",
    journal = "Astrophys. J. Lett.",
    volume = "961",
    number = "1",
    pages = "L10",
    year = "2024"
}

@article{Feng:2025ybf,
    author = "Feng, Wei-Xiang and Yu, Hai-Bo and Zhong, Yi-Ming",
    title = "{Black hole formation at low temperatures: Fermi degeneracy pressure}",
    eprint = "2510.24565",
    archivePrefix = "arXiv",
    primaryClass = "gr-qc",
    doi = "10.1103/g3df-sw3s",
    journal = "Phys. Rev. D",
    volume = "113",
    number = "6",
    pages = "063004",
    year = "2026"
}

@misc{SIDMcode,
  author = {Gu, Huapeng. and Jiang, Fangzhou. and Chen, Xian.},
  title = {{SIDM Halo Collapse Simulation in GR + Hydrodynamical}},
  year = {2026},
  publisher = {GitHub},
  journal = {GitHub repository},
  howpublished = {\url{https://github.com/Hua-Peng-G/SIDM}}
}

\end{document}